UNIVERSITY OF CALIFORNIA,
IRVINE

The Biophysical Micro-Environment's Influence on Cell Fate Decisions
During Macrophage Activation and Somatic Cell Reprogramming

DISSERTATION

submitted in partial satisfaction of the requirements
for the degree of

DOCTOR OF PHILOSOPHY

in Biomedical Engineering

by

Zachary Reitz

Dissertation Committee:
Assistant Professor Timothy L. Downing, Chair
Associate Professor Wendy Liu
Associate Professor Michelle Digman

2019

# DEDICATION

To

my father

for teaching me that science and empathy
must go hand in hand

and to my spouse

for giving me the foundation
I needed to succeed

# TABLE OF CONTENTS

# LIST OF FIGURES

# ACKNOWLEDGMENTS

I would like to acknowledge my committee chair, Dr. Timothy L. Downing for taking a chance on me. His drive and attention to detail helped form the foundation for the research presented in this dissertation.

I would like to thank my committee members, Dr. Wendy Liu and Dr. Michelle Digman for their support. Additionally, Dr. Liu's guidance and advice were essential for Section 1 of this dissertation.

I would also like to express gratitude to my peers both Praveen K. Veerasubramanian and Tri Andrew Quoc Phan. Their efforts were critical to the success and scale of the research presented here.

Furthermore, my spouse Stacey Reitz deserves infinite praise for supporting me through the process of putting together this dissertation.

I would also like to thank the funding sources that made this work possible including The National Science Foundation and The Simon's Foundation. The various facilities and organizations on campus that were critical to the research presented here include The Center for Multiscale Cell Fate, The Center for Complex Systems Biology, and the Edward's Lifesciences Center for Advanced Cardiovascular Technology.

# CURRICULUM VITAE

## Zachary Reitz

| | |
|---|---|
| 2014 | B.S. in Chemical Engineering, minor in Materials Engineering<br>California State Polytechnic University, Pomona |
| 2014-19 | Graduate Student Researcher, Biomedical Engineering<br>University of California, Irvine |
| 2015-19 | Teaching Assistant, Biomedical Engineering<br>University of California, Irvine |
| 2017 | M.S in Biomedical Engineering<br>University of California, Irvine |
| 2019 | Ph.D. in Biomedical Engineering,<br>University of California, Irvine |

## FIELD OF STUDY

The regulation of cell fate by epigenetics and the biophysical microenvironment

## PRESENTATIONS

Understanding the Role of the Adhesome in Somatic Cell Reprogramming UC Irvine Biomedical Engineering Seminar Series, February 2019

Epigenetic Consequences of Biomaterial Design on Macrophage Inflammatory Responses Biomedical Engineering Society Conference 2018, October 2018

The Adhesome as a Limiting Factor in iPSC Production MechBio Conference 2018: The Mechanome in Action, July 2018

Dynamic Adhesome Gene Regulation in Somatic Cell Reprogramming Edwards Lifesciences Cardiovascular Center Retreat, July 2018

The Epigenetic Consequences of Biomaterial Design UC Systemwide Bioengineering Symposium, June 2018

# ABSTRACT OF THE DISSERTATION

The Biophysical Micro-Environment's Influence on Cell Behavior
During Macrophage Inflammation and Somatic Cell Reprogramming

By

Zachary Reitz

Doctor of Philosophy in Biomedical Engineering

University of California, Irvine, 2019

Professor Timothy L. Downing, Chair

Cells are known to sense and respond to their mechanical microenvironment in profound ways. Various evidence has implicated the adhesome, a body of adhesion associated proteins, in several cell fate decisions, including differentiation and proliferation. Furthermore, recent work has demonstrated that substrate topography can modulate the epigenetic state of fibroblasts, facilitate cell reprogramming, and temper inflammatory activation. However, the effectors responsible for such phenomena remain poorly understood. Here we explore the effect of biomaterial design, adhesion, and intracellular forces of the epigenetic and transcriptional regulation of cell state during macrophage activation and somatic cell reprogramming.

Regarding macrophage activation, we relied entirely on murine bone marrow-derived macrophages (BMDMs). To study the effect of adhesion on BMDM behavior, we utilized microcontact printing to produce fibronectin patterned substrates ontop of which we seeded BMDMs. Next, we induced macrophage activation using lipopolysaccharide (LPS) and interferon-γ (INFγ) and found patterned substrates reduce inflammatory gene expression and histone-3 acetylation (H3Ac) while also increasing cellular elongation. We incorporated these results into a bioinformatic reanalysis of transcriptional and epigenetic data derived from a study of H3Ac protein binding during macrophage polarization. This analysis allowed us to identify epigenetic mechanisms linking the adhesome and inflammatory gene expression.

In our investigation of somatic cell reprogramming, we found extracellular matrix binding (ECM) and mechanosensitive ion channel activation reduce reprogramming efficiency. In addition to these results, we discovered that 104 adhesome genes are dynamically regulated during the reprogramming process. Subsequently, we performed an shRNA knockdown of each of these genes and found over 90% of our knockdowns increased the number of TRA-1-60 positive pluripotent colonies. Our knockdown of SHROOM3, a gene associated with apical cellular constriction, increased reprogramming efficiency by 27-fold. Further investigation into SHROOM3 identified a mechano-sensitive critical state transition, which may be necessary for successful reprogramming.

These observations establish adhesome gene expression and mechanical signaling as influential regulators of cell fate during macrophage activation and reprogramming. Moreover, our findings may guide future attempts to regulate cell state and fate using biophysical stimuli. They may also help shed light on cell behavior in various disease states involving cancer and inflammation.

# INTRODUCTION

Bioengineering is dependent on our ability to successfully regulate and control cell state and fate. Approaches to solving this problem often rely on the use of chemical factors or genetic modification. Anti-inflammatory drugs or coatings are frequently used to prevent fibrosis surrounding subcutaneous implants. However, these methods often engender undesired side effects and do not adequately mitigate the innate immune response.[1] Furthermore, previous work has found that somatic cell reprogramming, a process dependent on the exogenous expression of the Yamanaka factors, can be enhanced through the use of epigenetic enzyme inhibitors or by modulating growth factor concentration.[2,3] However, there remain poorly understood stochastic barriers to successful reprogramming.[4] Ultimately, future bioengineering strategies need to consider the biophysical microenvironment in addition to well known biochemical pathways.

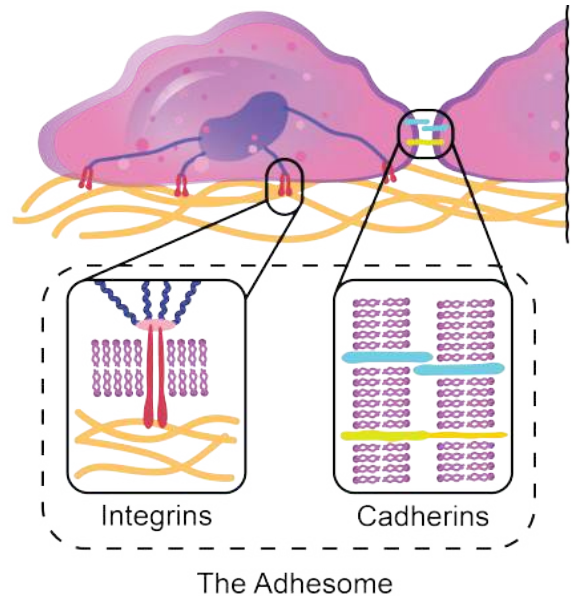

*Figure 1: The Adhesome is composed on integrins, cadherins, and all associated cytoskeletal or signalling proteins*

An extensive body of evidence exists to support the claim that mechanical cues are a critical regular of cellular behavior. Previous work has shown that engineered microscale surface topographies can lessen the activation of proinflammatory macrophages.[5–7] Similarly, such topographical cues are known to increase the efficiency of somatic cell reprogramming.[8] Using substrates with very low stiffness also attenuates macrophage-mediated inflammation and increases reprogramming efficiency.[9,10] Though these effects are observable, few efforts have conclusively deciphered the regulatory processes at play in these particular systems.

However, a variety of identified pathways may, in part, explain these inflammation and reprogramming related observations. First, micro-grooved surfaces can change global levels of epigenetic histone modifications, fundamental regulators of cellular identity, and their associated transferases.[8] Additionally, YAP/TAZ and MRTF-A/SRF are two well know transcriptional regulators that respond to mechanical cues and are critical for lineage commitment during cellular differentiation. [11,12] However, cellular adhesion and the actin cytoskeleton are necessary to activate these mechanosensitive pathways. As such, the adhesome, a body of genes representing all integrins, cadherins, and related proteins as seen in **Figure 1**, is a necessary prerequisite to mechanotransduction. [13,14] By studying the relationship between the biophysical microenvironment, mechanosensitive signaling pathways, and cell state, we aim to advance our ability to control cell fate in the contexts of macrophage inflammation and somatic cell reprogramming.

# SECTION 1

# The Biophysical & Epigenetic Regulation of Macrophage Activation

***Subsection 1.1: Micropatterned Surfaces Attenuate M1 Macrophage Activation***

Previous work established that micro-patterning material surfaces with fibronectin lines of widths as small as 20 μm will cause macrophages to express higher levels of the pro-healing marker Arginase. Conversely, such patterned surfaces appear to attenuate the protein level expression of inducable nitric oxide synthase (iNOS), a proinflammatory marker, in macrophages stimulated with the proinflammatory factors lipopolysaccharide (LPS) and interferon-gamma (INFγ).[7] Furthermore, work concerning both micropatterned surfaces and micro-grooved topographies found there is a positive correlation between line or groove width and iNOS expression.[5,7] Thus, we aimed to investigate the ability of micropatterned surfaces with lines as small as 5 μm to lessen the inflammatory response of macrophages activated with LPS and INFγ, also known as M1 macrophages. Furthermore, we looked to make a connection between inflammatory gene RNA expression, cellular elongation, and epigenetic markers of gene activation.

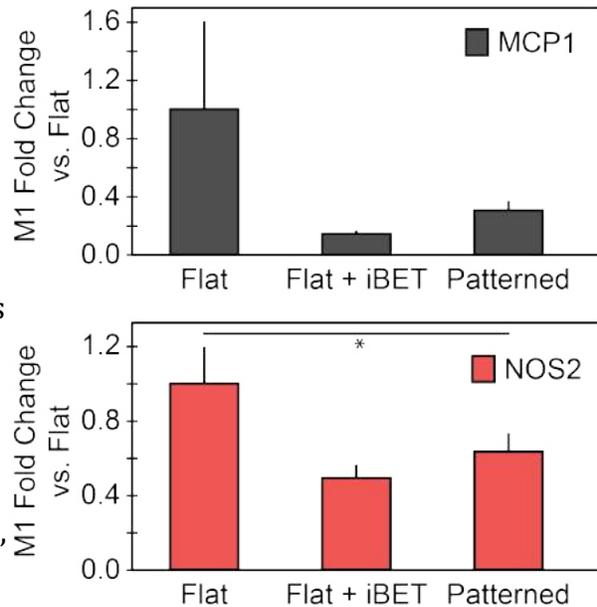

Figure 2: qPCR data for MCP1 (CCL2) or NOS2 (iNOS) normalized to flat condition. Error bars represent standard error. *P < 0.05

When we seed macrophages on patterned surfaces with 5 μm fibronectin lines and stimulate towards M1 activation, we find the patterned surfaces reduced the gene expression of iNOS by roughly forty percent. Furthermore, patterned surfaces also reduced the gene expression of monocyte chemoattractant protein1 (MCP1 also known as CCL2). Both observations are presented in **Figure 2**. From these results, we can conclude previous findings regarding the reduction of iNOS express on patterned surfaces is likely due to regulatory changes before and not after transcription.

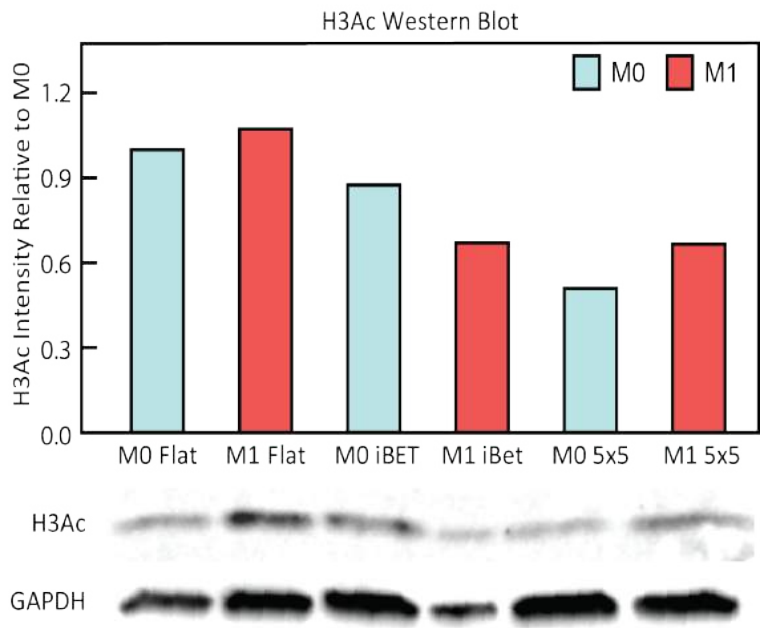

Figure 3: Western Blot for H3Ac Normalized to GAPDH including patterned (5 μm x 5 μm). Fold changes are relative to an M0 control

With that said, we looked to pre-transcriptional epigenetic regulatory mechanisms as a possible explanation for our observations. We know from previous work that histone three acetylation (H3Ac), a marker of gene activation, is regulated in other systems by biophysical cues.[8] Furthermore, histone deacetylase 3 (HDAC3) activity is involved in the downregulation of inflammatory gene expression when M1 macrophages are spatially confined.[15] Thus we decided to look at H3Ac levels in our M1 macrophages on patterned (5 μm by 5 μm) versus flat surfaces. We also included M0 unstimulated macrophages as a control. We found that while M1 stimulation increased global levels of H3Ac, culturing macrophages on patterned surfaces reduced H3Ac in M0 and M1 macrophages, as seen in Figure 3. Given H3Ac's role in gene activation, a reduction of global H3Ac indicates epigenetic regulatory mechanisms may explain the observed decrease in iNOS and MCP1 gene expression.

Previous work indicates bromodomain proteins (BRDs), known to be epigenetic readers of H3Ac, play a crucial role in inflammatory gene expression. Researchers were able to inhibit BRD protein binding using a synthetic mimic of histone acetylation known as iBET. This inhibition resulted in the suppression of proinflammatory M1 macrophage gene expression.[16] When we used iBET in our system, we found that BRD inhibition resulted in a decrease in iNOS and MCP1 expression (**Figure 2**), as we might expect. However, we also saw a modest decrease in H3Ac in our iBET treated conditions (**Figure 3**). This decrease may result from an interruption of proinflammatory feedback mechanisms that serve to increase global H3Ac during M1 activation.

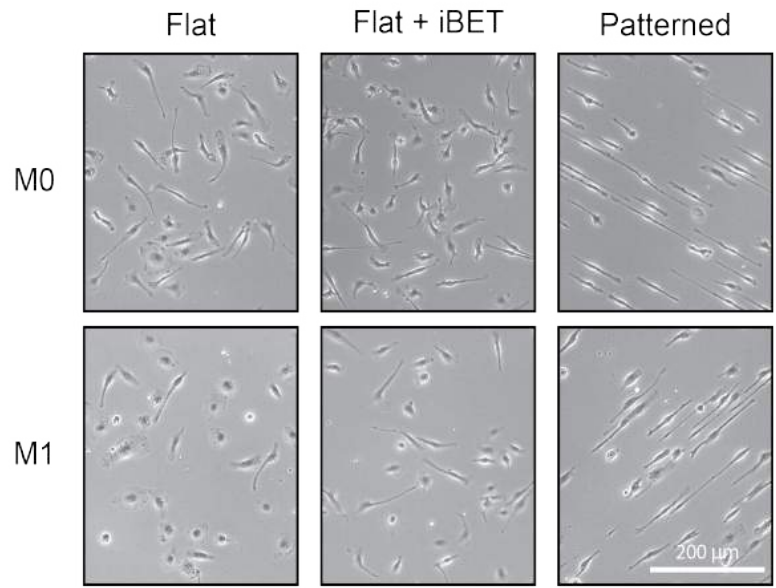

*Figure 4: Phase-contrast microscopy representative images of M0 and M1 macrophages with either iBET treatment or micropatterned surfaces*

Given the relationship between cell shape and M1 macrophage activation [7], we wanted to investigate the link between iBET treatment, micropatterns, and cellular elongation. We were able to image (**Figure 4**) and quantify the average cell shape and aspect ratio using phase-contrast microscopy. (**Figure 5**) As expected, the M1 macrophages are less spindle-like than the rounder M0 macrophages on a flat surface. Additionally, our results indicate a significant increase in cellular elongation occurs when M0 or M1 macrophages are seeded on a patterned surface relative to a flat surface. Intriguingly, iBET treatment resulted in M1 macrophages with a significantly higher cellular aspect ratio, akin to M0 macrophages on a flat surface. This observation likely means that epigenetic mechanisms are responsible for the synergistic relationship between M1 activation and cell shape. However, such synergy is likely dependent on transcriptional changes not identified by this experiment.

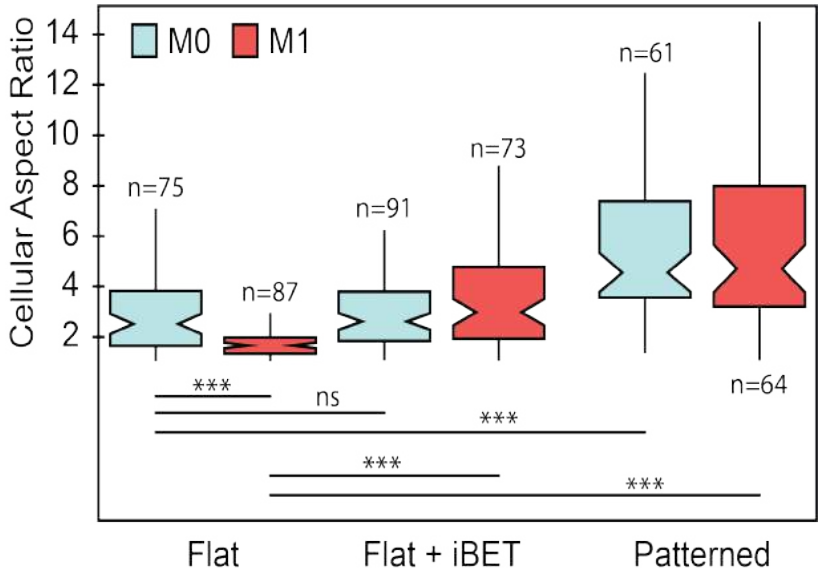

Figure 5: Cellular aspect ratio across all conditions. (Longest axis divided by maximum perpendicular width) $P^* < 0.05$ $P^{**} < 0.005$ $P^{***} < 0.001$

*Subsection 1.2:*

**Adhesome Gene Expression is Dynamic During Macrophage Activation**

Given the relationship between cell morphology and M1 macrophage activation, we were interested in investigating the role adhesion proteins play during macrophage differentiation and polarization. With that said, we applied a bioinformatic approach to reanalyze and visualize findings from previous data sets. **Figure 6** shows adhesome gene expression during the differentiation of HL-60 myeloid progenitors into macrophages. It also shows adhesome gene expression during LPS stimulation from the same dataset.[17] What is clear from this visualization is that adhesome gene expression is highly dynamic during macrophage differentiation and, eventually, M1 activation. This data indicates that adhesome genes may be responsible for some of the morphological and functional changes characterized by M1 activation. Additionally, these gene expression dynamics suggest that parts of the adhesome may play a regulatory role during M1 activation and differentiation.

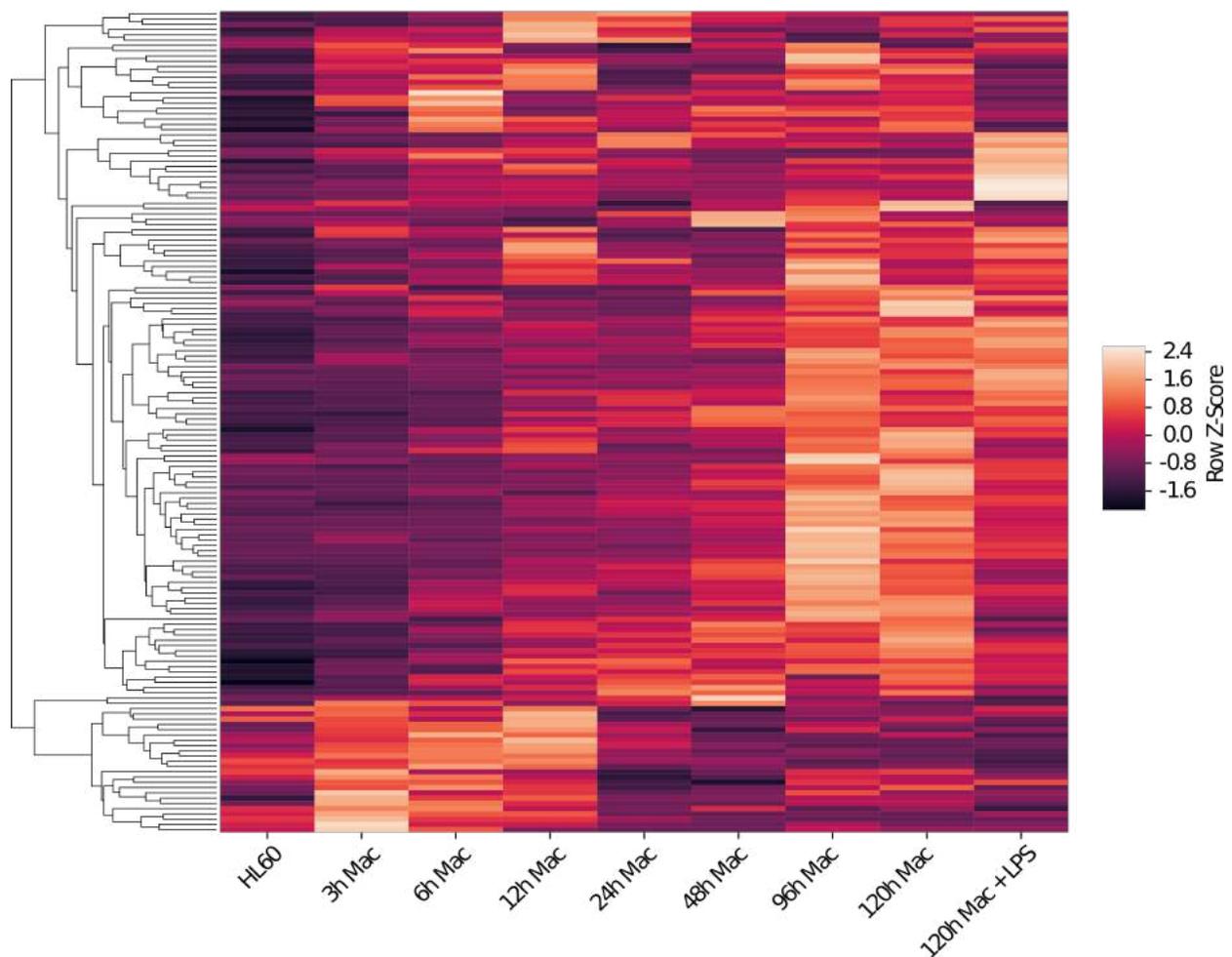

*Figure 6: Heatmap of adhesome gene expression during macrophage differentiation and LPS stimulation.*

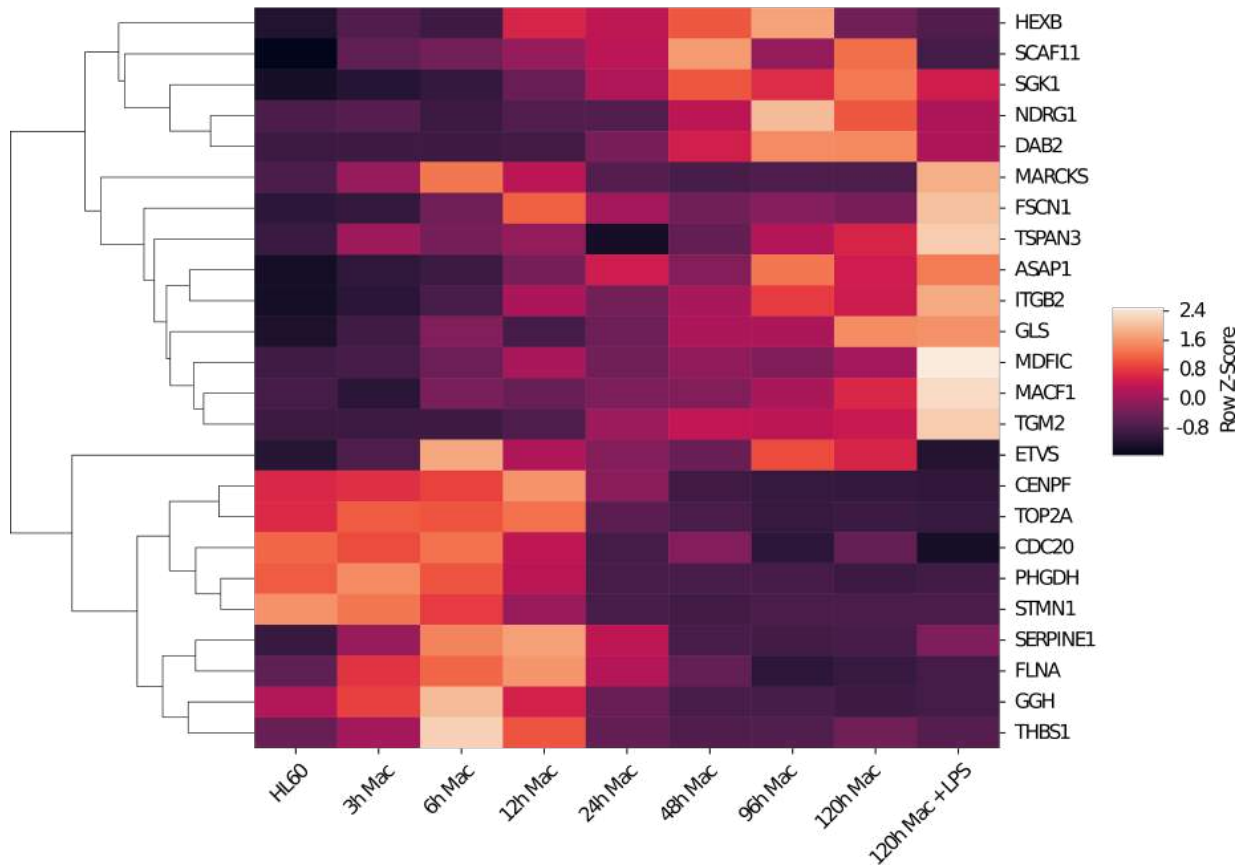

*Figure 7: Heatmap of dynamic YAP/TAZ target gene expression during macrophage differentiation and M1 activation*

However, mechanosensitive pathways are necessary in order to trigger the previously observed shifts in epigenetic markers such as H3Ac and proinflammatory gene expression. Previous work has implicated the myocardin related transcription factor A and serum response factor (MRTF-A/SRF) mechanosensitive pathway as a critical regulator of M1 activation.[15] However, yes-associated protein and transcriptional coactivator with PDZ-binding motif (YAP/TAZ) form another well known mechanosensitive pathway responsible for lineage commitment in mesenchymal stem cells.[11] With that said, we decided to investigate the role of YAP/TAZ target gene expression in the data set previously described.[17] Intriguingly, YAP/TAZ target gene expression undergoes dynamics similar to those experienced by the adhesome. (**Figure 7**) Furthermore, LPS stimulates results in the upregulation of several YAP/TAZ targets. Given the mechanical changes that occur during M1 activation, YAP/TAZ target upregulation is a good indication that such mechanical cues result in transcriptional changes as well.

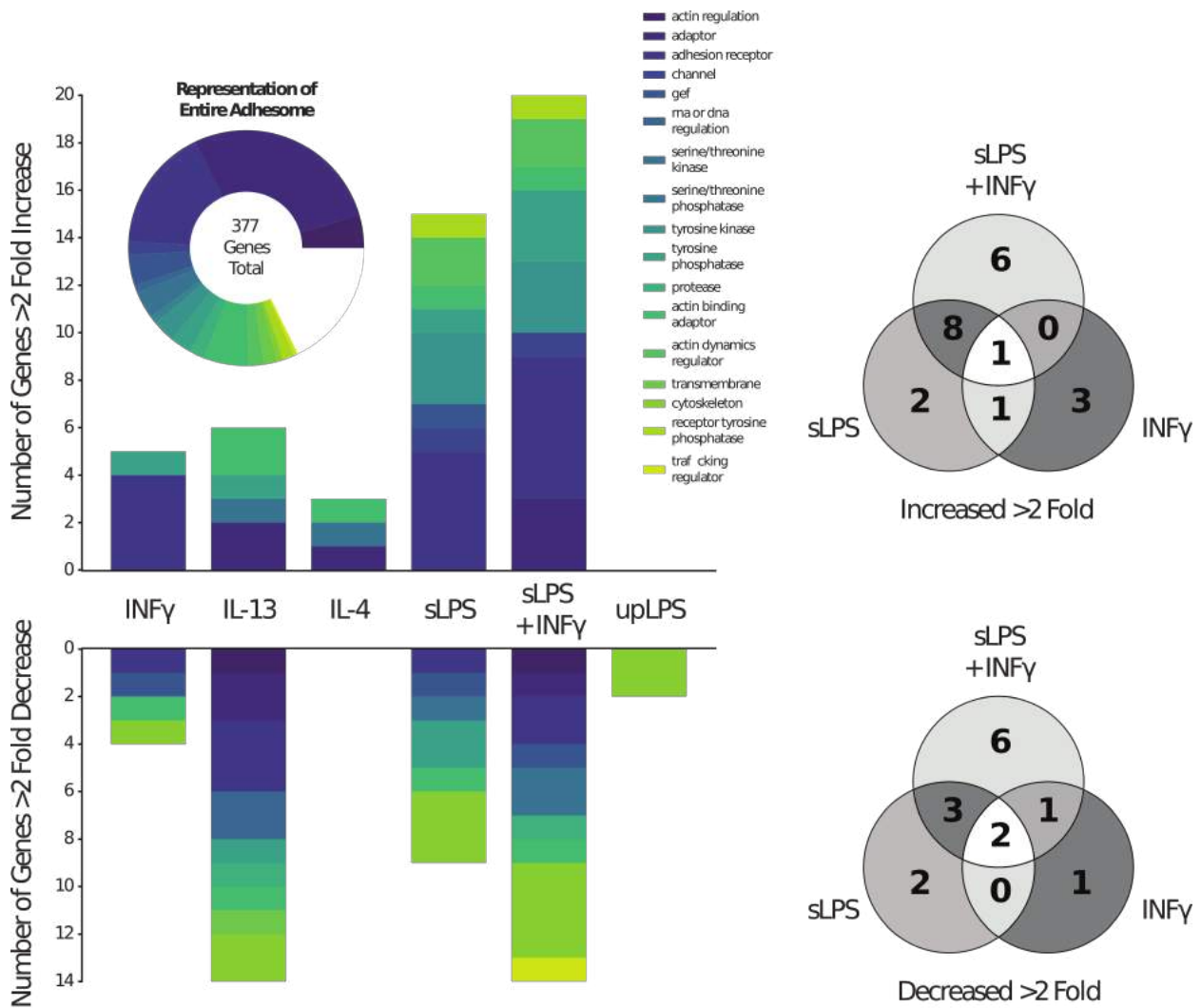

*Figure 8: Adhesome gene regulation is dependent on macrophage stimulation conditions. Venn diagrams represent the overlap in differentially regulated genes.*

Nevertheless, M1 activation is a catch-all term for a variety of proinflammatory signaling pathways. Cytokines such as INFγ and factors like LPS engender distinct inflammatory responses in macrophages.[18] With that said, we set out to reanalyze transcriptomic data to see how various chemical M1 activators result in adhesome gene expression dynamics. (**Figure 8**) From our analysis, we found both standard LPS (sLPS) and INFγ differentially regulated unique gene sets within the adhesome, relative to controls. Furthermore, INFγ and sLPS differentially regulate a unique set of genes when used together as opposed to independently. Interestingly, ultra-pure LPS (upLPS), which activates a more limited set of surface receptors than sLPS, has a minimal effect on adhesome gene expression. These findings suggest that M1 macrophage activation may regulate the adhesome and incorporate mechanical signaling in a variety of independent pathways.

**Subsection 1.3:**

**Inflammatory and Adhesome Genes are Co-regulated During M1 Activation**

In order to get a better understanding of the inflammatory gene suppression that occurs when M1 macrophages are cultured on micropatterned surfaces, we utilized a Nanostring inflammatory panel. From our results, we can ascertain that patterned surfaces modulate inflammatory gene expression after as little as 2 hours of stimulation with INFγ and LPS. However, such differential expression is more apparent after 16 hours of stimulation, where there is a significant downregulation of iNOS and MCP1. (**Figure 9**) However, unlike previous studies regarding spatial confinement, we find patterned surfaces upregulate the gene expression of Nfκb, a proinflammatory transcription factor.[15] This observation supports the idea that relationships between mechanical cues, the adhesome, and M1 activation embody a network of behaviors and can not be easily summarized.

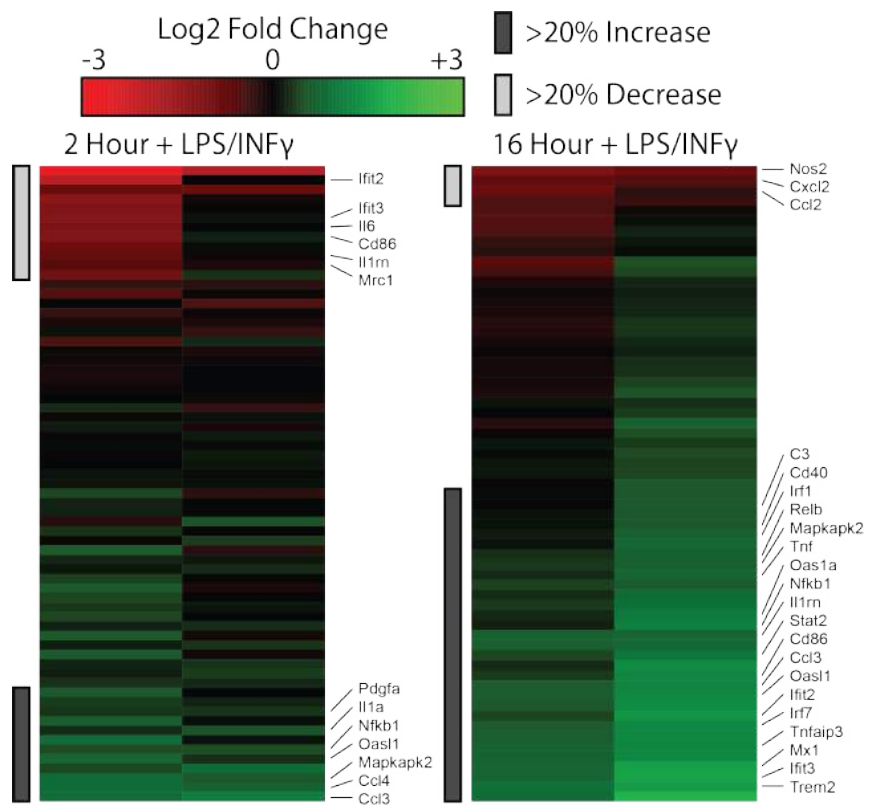

*Figure 9: Inflammatory genes are differentially regulated by micropatterned surfaces. Color-scale represents fold change relative to flat surfaces.*

With that in mind, we compared these differential regulated genes to the dynamics observed when M1 macrophages undergo iBET treatment.[16] Although there was some overlap in the gene sets down-regulated by micropatterned surfaces and iBET treatment, very few genes upregulated by the patterned surfaces experienced similar changes during iBET treatment. (**Figure 10**) Such an observation is not unlikely, as iBET globally targets readers of H3Ac, a marker for gene activation, while the pathways affected by patterned surfaces may have more specific targets.

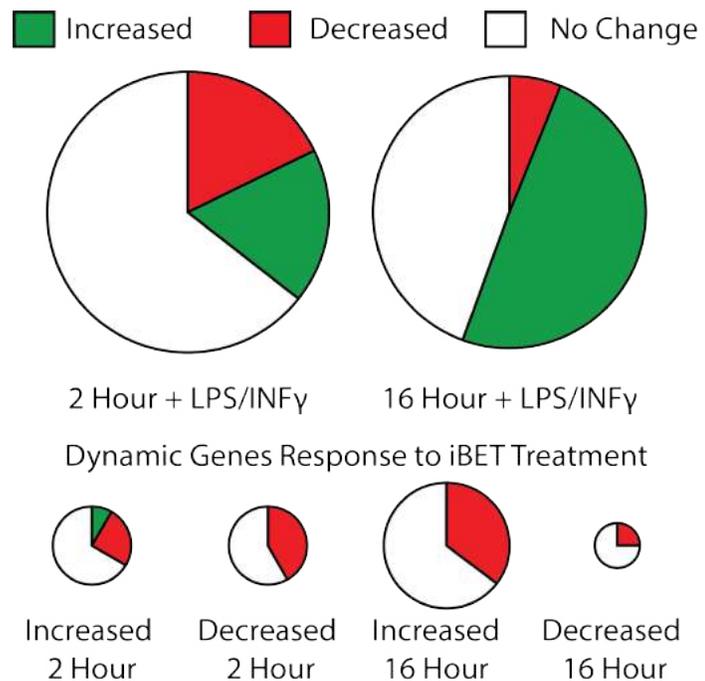

Figure 10: Dynamic inflammatory gene expression on patterned relative to flat surfaces. Differentially regulated genes are then compared to expression changes resulting from iBET treatment.

To decipher these nuanced systems of gene regulation, we leveraged the previously mentioned data set to build a gene expression correlation network. The network represented in **Figure 11** was build using adhesome genes and the inflammatory genes that were differentially regulated by patterned surfaces. Within the network, there is a notable and multifaceted relationship between adhesome and inflammatory gene expression. Furthermore, iNOS, MCP1, and Nfκb all exist within separate parts of the network. This lack of a correlative relationship suggests these genes activated as part of separate and not opposing pathways, as the inflammatory panel might suggest. With that said, genes upregulated along with Nfkb, as part of the response to micropatterned surfaces, are tightly networked with Nfκb. With that in mind, we built a second gene correlation network using StringDB. (**Figure 12**) Taken together, our independently generated network and the StringDB network indicate that the relationships between genes within our network are not artificial. As such, the newly identified relationships between adhesome and inflammatory genes likely have biological meaning.

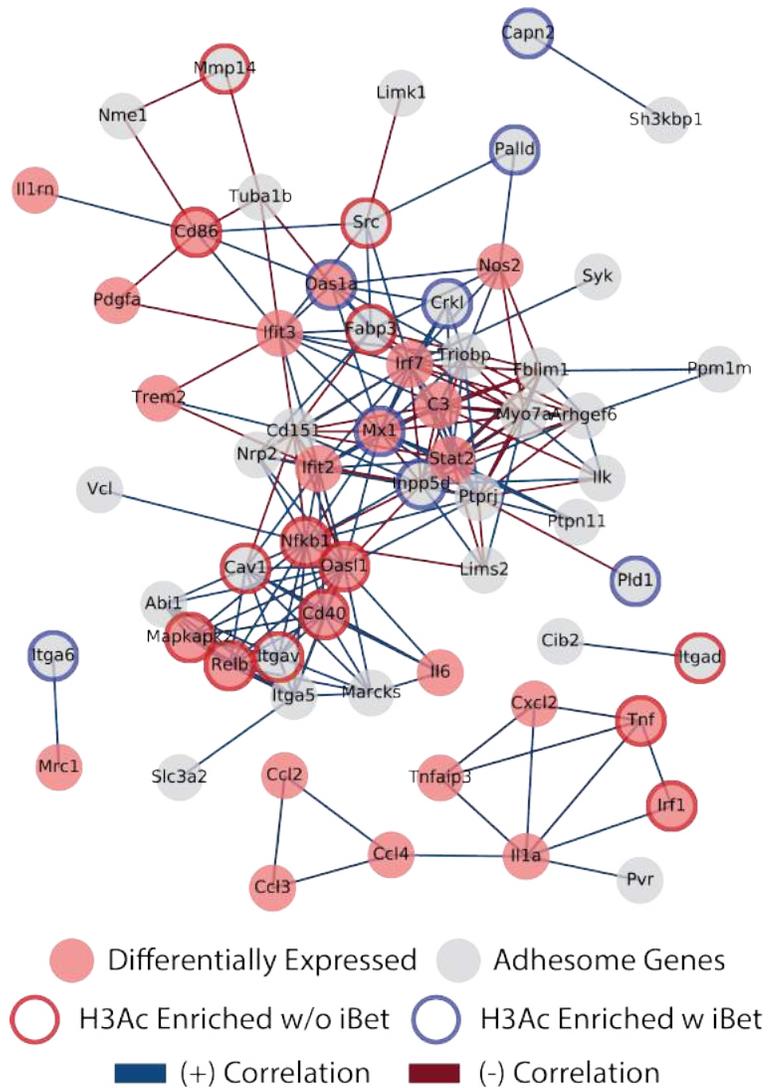

Figure 11: Gene correlation network built using inflammatory genes regulated by micropatterned surfaces and adhesome genes in macrophages.

Following the creation of these gene correlation networks, we sought to identify the regulatory mechanisms that result in such gene co-regulation. To accomplish this goal, we used a differential enrichment analysis method on a data set of ChIP-seq H3Ac tracks for macrophages stimulated with LPS, with or without iBET treatment.[16] From this analysis, we can identify genes where H3Ac increased during LPS stimulation without iBET, but where H3Ac does not increase with iBET. (**Figure 13**) Intriguingly, we found a closely connected region of the gene expression correlation network contains adhesome and inflammatory genes that have such epigenetic shifts in response to LPS and iBET treatment. This cluster includes both Nfκb and adhesome genes such as Cav1 and ItgaV. From these observations, we can conclude that the relationship between mechanical signaling, inflammatory gene expression, and adhesome gene expression may be dependent on epigenetic mechanisms.

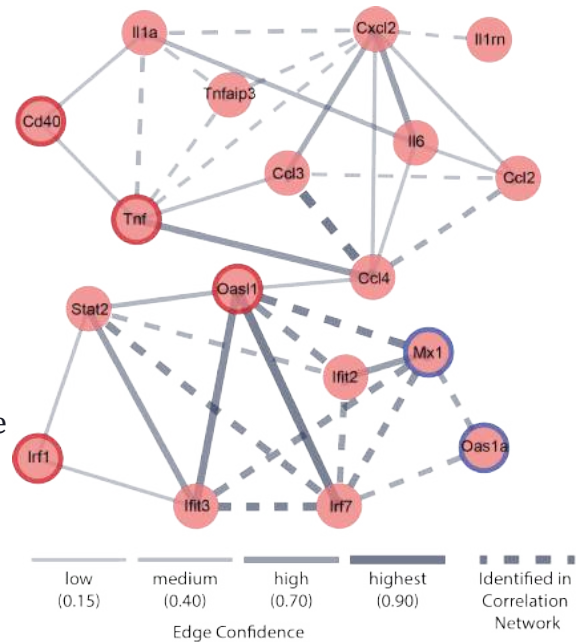

Figure 12: StringDB gene correlation network built using genes from the network in Figure 11

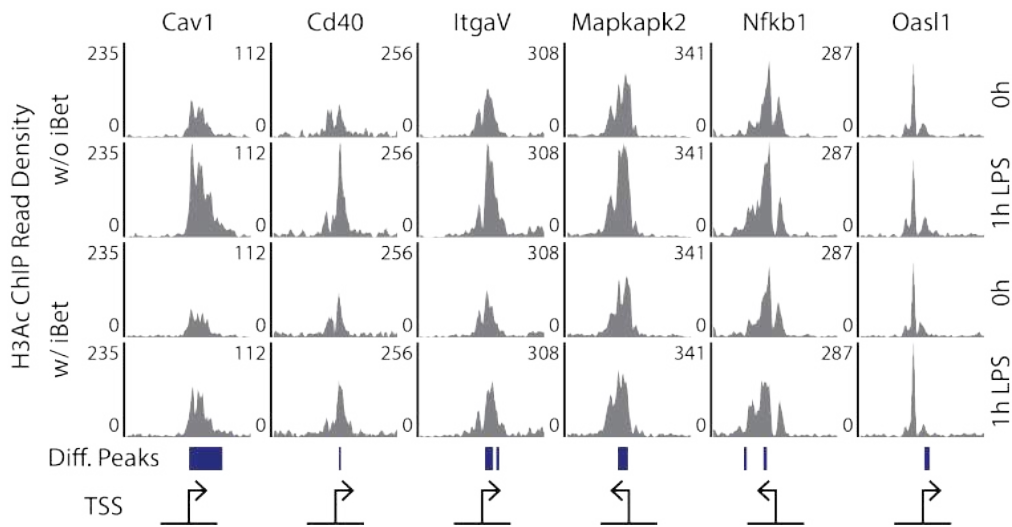

Figure 13: ChIP-seq tracks for H3Ac enrichment in the promoters of several closely genes. Ranges surrounding the transcriptional start site are +/- 5kb. Regions where H3Ac is increased during LPS stimulation, only without iBET are labeled in blue.

## Section 1.4:

### The Relationship Between Mechanical Cues, The Cytoskeleton, and H3Ac

After noting that H3Ac decreased in BMDMs that are cultured on a micropatterned surface, we aimed to validate these findings through an orthogonal approach. With that in mind, we used immunostaining to quantify histone acetylation within cellular nuclei. (**Figure 14**) The results from this experiment mirror the results of the previous western blot. Here, we show that while M1 stimulation increases nuclear H3Ac, micropatterned surfaces decrease H3Ac in both M0 and M1 macrophages. Furthermore, these observations are statistically significant due to minimal cell to cell heterogeneity.

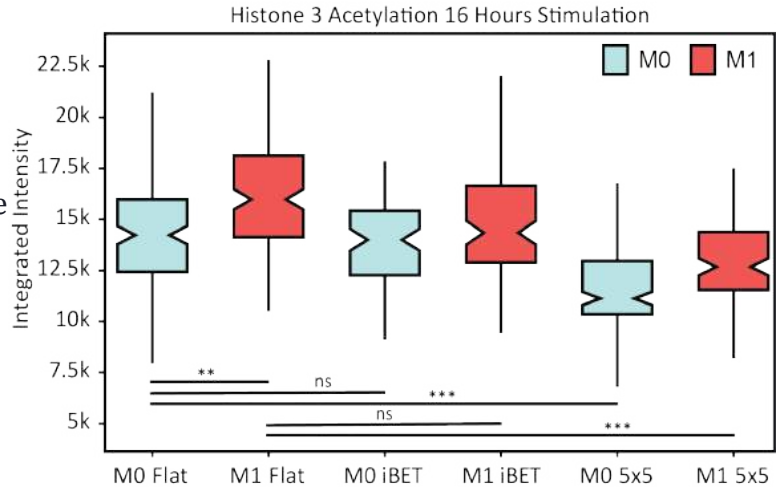

Figure 14: Nuclear H3Ac in macrophages after 16 hours of stimulation. Conditions include M0 and M1 activation states as well as either iBET treatment or micropatterned substrates, $P^* < 0.05$ $P^{**} < 0.005$ $P^{***} < 0.001$

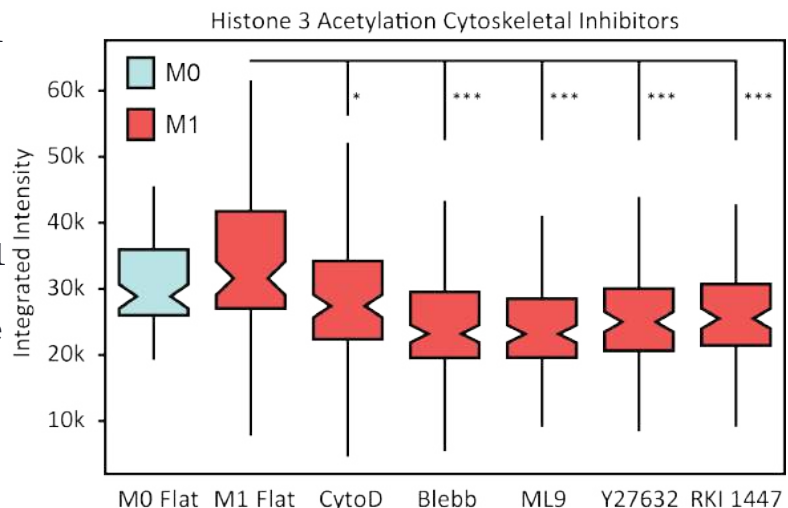

Figure 15: Nuclear H3Ac in macrophages after 16 hours of stimulation. Conditions include M0 and M1 activation states as well as either CytoD, Blebb, ML9, Y27632, or RKI-1447 treatment, $P^* < 0.05$ $P^{**} < 0.005$ $P^{***} < 0.001$

Given the relationship between H3Ac regulation and the actin polymerization in macrophages, we aimed to study the role of cytoskeletal reorganization in nuclear H3Ac. [15] Therefore, we employed a variety of cytoskeletal inhibitors, including cytochalasin D (Cyto-D), myosin light chain kinase inhibitor (ML-9), blebbistatin, and the Rho kinase (ROCK) inhibitors RKI-1447 and Y-27632. Using immunostaining, we found that every one of the cytoskeletal inhibitors significantly reduced nuclear H3Ac in M1 macrophages. (**Figure 15**) Such an observation suggests that micropatterned surfaces reduce H3Ac by altering cytoskeletal organization in macrophages.

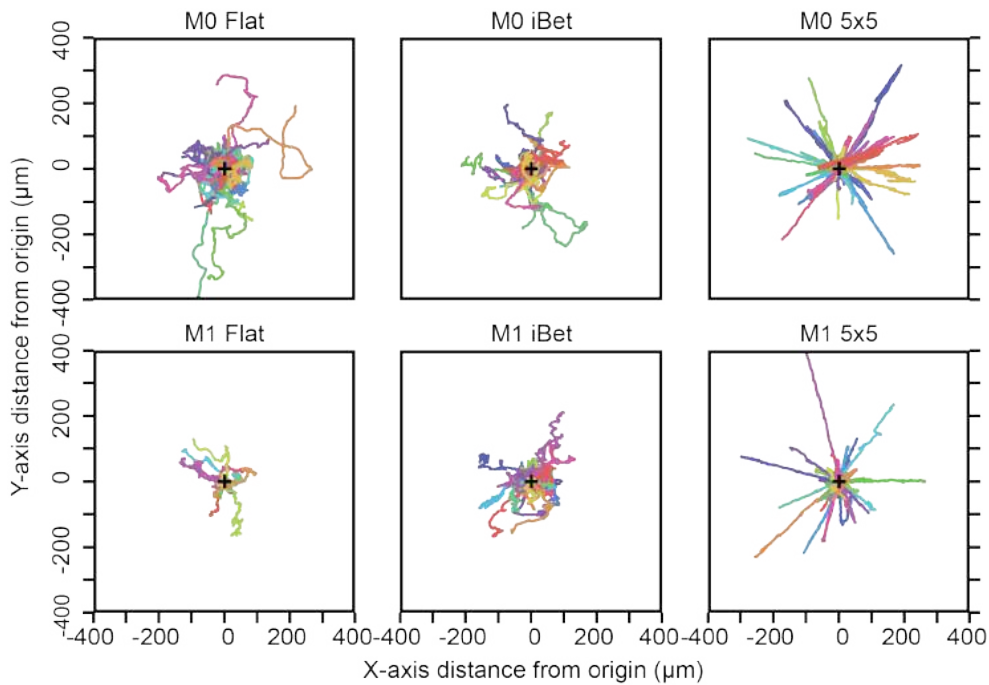

*Figure 16: Cellular migration tracks for M0 and M1 macrophages with either iBET treatment or micropatterned surfaces*

Furthermore, previous findings have implicated cytoskeletal organization in the migratory capacity of macrophage and their ability to produce force. [19] With that said, we aimed to investigate the relationship between macrophage motility, polarization, epigenetic regulation, and micropatterned surfaces. By tracking the centroid of various M0 and M1 macrophages, we can see that M0 macrophages appear to be highly motile, whereas M1 macrophages do not venture far from their original position. (**Figure 16**) We subsequently quantified the cellular motility of macrophages in each condition by measuring their average velocity and maximum displacement over the tracking period. In **Figures 17** and **18**, we show that while the velocity and maximum displacement of M1 macrophages are less than that of M0 macrophages, iBET treatment and micropatterning significantly increases those metrics in M1 macrophages. These observations further corroborate the observed epigenetic relationship between the cytoskeleton, adhesion, and M1 macrophage activation.

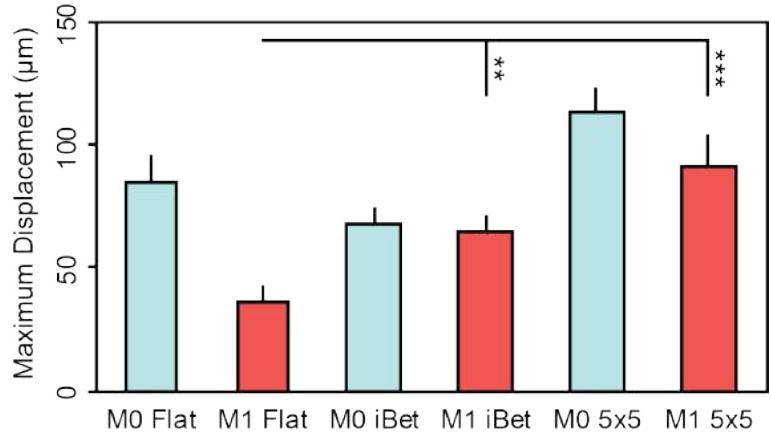

Figure 17: Maximum displacement of M0 and M1 macrophages with either iBET treatment or micropatterned surfaces, P* < 0.05 P** < 0.005 P*** < 0.001

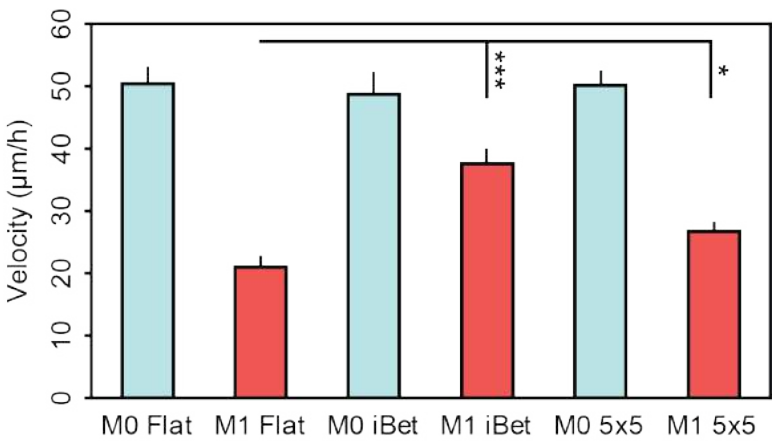

Figure 18: Average velocity of M0 and M1 macrophages with either iBET treatment or micropatterned surfaces, P* < 0.05 P** < 0.005 P*** < 0.001

# SECTION 2

## Cell-generated force and adhesion contributes to a bottleneck during cell reprogramming

***Subsection 2.1:***

***Adhesion and Mechano-Signaling are closely Tied to Successful Reprogramming***

The extracellular environment is known to influence critical cell fate decisions during differentiation, metastasis, and innate immunity[20–22]. Recently, many works have illustrated the importance of material properties on cellular reprogramming. These properties, such as decreased stiffness, micro-grooved topographies, and even 3D material matrices, have been shown to improve reprogramming efficiency[8,10,23]. It seems likely that the enhanced reprogramming observed in these scenarios is a result of adhesion mediated signaling. Such adhesion is often a result of integrin to ECM binding, which is dependent on integrin-binding site availability in the ECM. Here we posit that the protein composition of the extracellular matrix plays an impactful role in the reprogramming process.

To test this hypothesis, we coated polystyrene tissue culture plastic with various ECMs such as fibronectin, type I collagen, gelatin (hydrolyzed collagen), and two commercially available cancer-derived ECMs known as Matrigel® and Geltrex™. The latter two represent a more chemically complex and thus more biologically relevant ECMs. We then seeded and reprogrammed immortalized human fibroblasts (hiF-Ts), that contain a doxycycline-inducible OKSM (Oct4, Klf4, Sox2, c-Myc) gene cassette, to become induced pluripotent stem cells (iPSCs) as previously described by Cacchiarelli et al.[2] After, 12 days of reprogramming, the number of pluripotent stem cell colonies, as measured by TRA-1-60

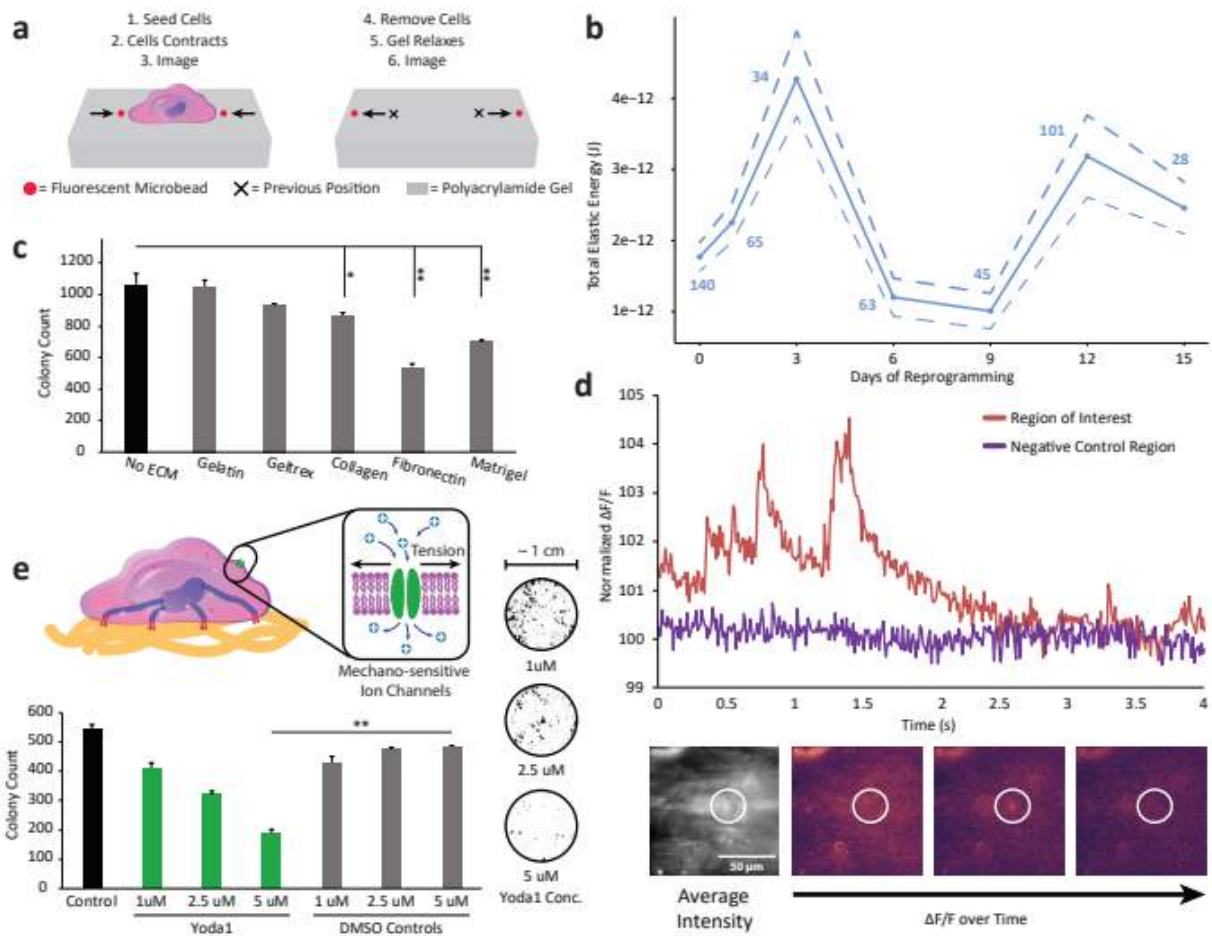

*Figure 19: Adhesion and force generation are dynamic during and an impediment to reprogramming. (a) Depiction of traction force microscopy (TFM)to measure cell generated forces (b) Traction forces are dynamic during the process of reprogramming (c) ECM coating reduces reprogramming efficiency (d) Calcium signalling, indicative of mechanosensitive Piezo1 channel activation, is active on day 12 of reprogramming (e) Treatment with Yoda1, a Piezo1 agonist, decreases reprogramming efficiency, $P^* < 0.05$ $P^{**} < 0.005$ $P^{***} < 0.001$*

staining, were counted and compared (**Figure 19c**). Our results indicate that fibronectin coating reduced reprogramming efficiency the most with a ~49% decrease in pluripotent colonies. Matrigel followed with a 39% reduction in colony count with collagen type I and Geltrex had less of an effect, 19% and 10% respectively. Gelatin had a negligible impact with a ~2% reduction in overall efficiency. This evidence suggests integrin binding may be a critical inhibitor of the reprogramming process.

Previous work has shown that cellular adhesion differs between terminally differentiated cells, partially reprogrammed cells, and iPSCs[24]. Additionally, signaling downstream of cell to ECM interactions and intracellular generated traction forces are known to direct cell fate decisions and transitions during differentiation[25]. Thus, we set out to measure traction force generation and signaling during the reprogramming process. Using traction force microscopy (TFM), as previously described[26], we found hiF-Ts in the early stages of reprogramming abate traction force generation when represented as total elastic energy (**Figure 19b**). Later during the reprogramming process, we observe an increase in cellular traction forces before the completion of the reprogramming process. Respectively, an early loss of somatic cell identity and subsequent transient activation of differentiation pathways, as previously observed[2], may partially explain such dynamic force generation.

Additionally, mechano-sensitive pathways, including stretch-activated channels such as Piezo1, have been shown to play a role in neural stem cell lineage commitment[27] and thus may counteract the reprogramming process. To test the potential for such a possibility, we first used total internal reflectance fluorescence microscopy to confirm that Piezo1 channel activation does indeed occur naturally in reprogramming hiF-Ts, as represented in (**Figure 19d**). This observation indicates that mechanically induced ion channel signaling is occurring during the reprogramming process. Next, we treated reprogramming hiF-Ts with Yoda1, a small molecule drug previously shown to specifically activate Piezo1 sans mechanical stimuli[28]. As can be seen in (**Figure 19e**), Yoda1 treatment ultimately reduces reprogramming efficiency when compared to controls. This finding suggests that mechano-signaling via mechano-sensitive ion channels may indeed act as a barrier to reprogramming.

***Section 2.2: Adhesome Gene Expression is Dynamic During Reprogramming***

Since ECM adhesion and cellular force generation are likely inducing pathways in opposition to somatic cell reprogramming, we wanted to elucidate the role of the proteins and genes responsible. As such, the Adhesome, as defined by Zaidel-Bar et al[13,14] and which contains integrins, cadherins, and other associated proteins, appeared to be a prudent target for investigation. A large body of work has previously implicated integrins and cadherins as crucial drivers of developmental pathways and processes such as Wnt3 signaling, EMT, neural tube closure, and tissue morphogenesis[29]. In particular, the EMT transition during neural tube closure, resulting in migratory mesenchymal-like neural crest cells, requires the dynamic expression of E-cadherin and N-cadherin[30]. The regulatory capacity for such adhesion proteins and their associated signaling pathways leads us to believe they likely impair the reversion of developmental processes that occur during reprogramming.

To obtain a broader understanding of the Adhesome's role during reprogramming, we reanalyzed a hiF-T reprogramming RNA-seq timeline, generated by Cacchiarelli et al. [2]. We aimed to identify adhesome genes whose expression profiles during reprogramming were dynamic, defined as a greater than six-fold change with expression levels higher than five fragments per kilobase million (FPKM). From our analysis we discovered that 105 Adhesome genes had dynamic expression patterns during reprogramming. (**Figure 20b**). Notably, a large proportion of these dynamically expressed adhesome genes are down-regulated early in the reprogramming process (**Figure 20c**). The remaining set of genes is transiently or permanently up-regulated upon arrival at a pluripotent state. Interestingly, these observations appear to correspond with the dynamic traction forces observed earlier. This finding suggests an association exists between dynamic adhesome gene expression, intracellular generated forced, and the sequential loss of somatic identity and transient up-regulation of differentiation pathways noted in previous work[2].

      To test the possible regulatory effect of each of these dynamic genes on reprogramming, we designed three shRNAs to target and ultimately reduce the mRNA expression of each gene. Every shRNA expressing construct was separately transduced into hiF-Ts using a lentiviral vector, before the start of reprogramming. Upon completion of the reprogramming process at day 24 (**Figure 20d**), we measured the effect of each shRNA on overall reprogramming efficiency relative to non-targeting shRNA controls. As we observed in (**Figure 20e**), a vast majority of our shRNA knockdowns (kd) drastically improved overall reprogramming efficiency. Given that ROCK inhibitor (Y-27632 or ROCKi) and lysine-specific histone demethylase one inhibitor (LSD1i) are also known to enhance reprogramming efficiency[2], we wanted to know if Adhesome gene knockdowns achieve improvements when combined with the effects of these small molecules. As such, we conducted the same shRNA screen as before and added LSD1i and ROCKi to the reprogramming cell culture media. Upon completion of reprogramming on day 16, which occurs earlier due to LSD1i's propensity to accelerate the reprogramming process, we measured overall reprogramming efficiency. Here we noted the vast majority of shRNA Adhesome gene knockdowns again improved reprogramming efficiency, though the magnitude of these improvements was somewhat lesser than without ROCKi and LSD1i. These findings, taken together, suggest that cellular adhesion and signaling plays a critical

role in the process of reprogramming. However, this evidence alone does not in itself confirm the possibility that these adhesome genes are responsible for facilitating differentiation signaling pathways that may run counter to somatic cell reprogramming.

Intending to get a better understanding of how adhesome genes are co-regulated during reprogramming, we built weighted gene correlation networks using adhesome gene expression data from our own reprogramming RNA-seq experiment. (**Figure 20f**) From these networks, we can see a high correlation between a wide variety of adhesome genes across days 6, 9, 12, and 15 of reprogramming. Intriguingly, the network appears most dense on day 12. This observation suggests that a transient event or state within the reprogramming process is intrinsically linked to cell adhesion by the genes in the adhesome. Furthermore, this link is likely not one-directional, given the increase in reprogramming efficiency caused by adhesome shRNA knockdowns. However, the signaling pathways associated with the most successful of these knockdowns can not be identified through these networks alone.

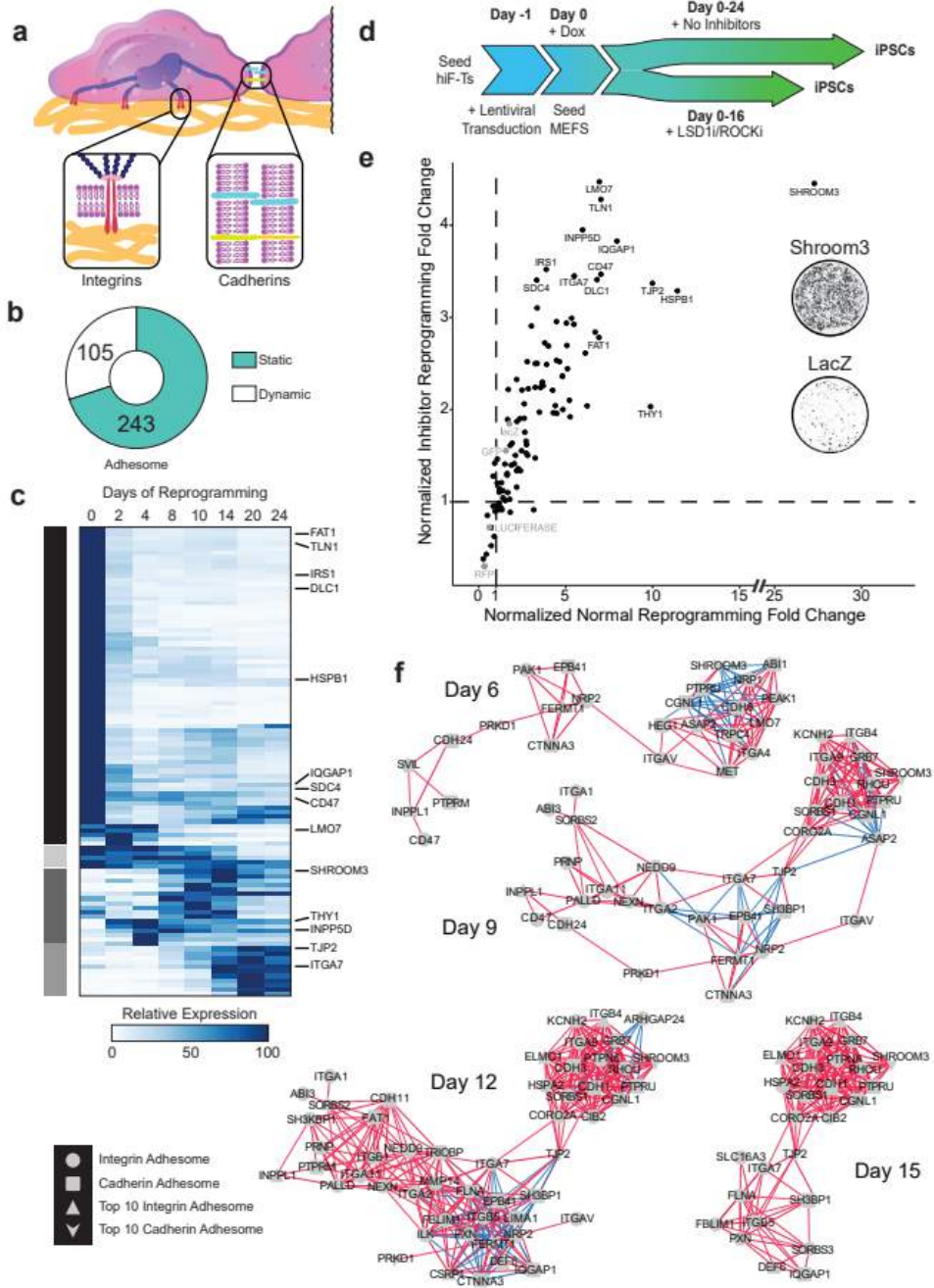

*Figure 20: Adhesome gene expression is dynamically regulated during reprogramming and is an impediment to successful reprogramming. (a) Diagram of the adhesome containing integrins, cadherins, and associated proteins (b) Fraction of adhesome genes that are dynamic, i.e. experiences greater that 6-fold change over the course of reprogramming (c) Heatmap of dynamically expressed adhesome genes clustered by expression profile (d) shRNA knockdown experimental timeline (e) Pluripotent TRA-1-60 colony counts relative to control conditions for all shRNA knockdowns of dynamic adhesome genes with or without LSD1i and ROCKi (f) Weighted gene correlation networks for day 6, 9, 12, and 15 of reprogramming built using integrin or cadherin related adhesome genes.*

***Section 2.3: Knockdown of SHROOM3 Differentially Regulates Mechanosensitive and Developmental Pathways***

From our screen of the dynamic adhesome, SHROOM3, in particular, stands out having increased reprogramming efficiency by 27 fold without inhibitors. Additionally, SHROOM3 is transiently expressed during reprogramming. Such dynamics suggests it may be playing a role in intermediate cell fate transitions on the path to pluripotency. Previous evidence has indicated SHROOM3 plays a vital role in the apical constriction of epithelial cells [31], a process necessary for developmental morphogenesis, neural tube closure, and subsequent EMT of neural crest cells[30]. Furthermore, SHROOM3 coordinates with and activates RhoA and ROCK1/2 to initiate myosin contraction within the cellular apex[32]. As such, we aimed to identify the gene regulatory networks that might, in part, explain the large improvement in reprogramming efficiency observed in our SHROOM3 kd.

To accomplish this goal, we produced an RNA-seq reprogramming timeline on our SHROOM3 kd and controls with samples every three days until day 15. In this way, we were able to capture transient changes in gene expression specific to the SHROOM3 kd condition (**Figure 21a**). Using gene set enrichment analysis, we identified a variety of enriched ontologies that as a whole appear to relate to late morphogenesis and transiently unregulated differentiation pathways previous work has described (**Figure 21d**).[2] These ontologies also depict what appear to be broad changes in gene regulation related to cellular adhesion and ECM remodeling. Previous work suggests there may be a link between these observed differences in gene regulation of the ECM and developmental/cancer-associated cell state transitions such as EMT.[33] However, these observations do not in and of themselves give us a clear idea as to what mechanisms may be driving such changes. Nor do they give us a concise perspective as to the manner in which such mechanisms are being perturbed by our kd of SHROOM3.

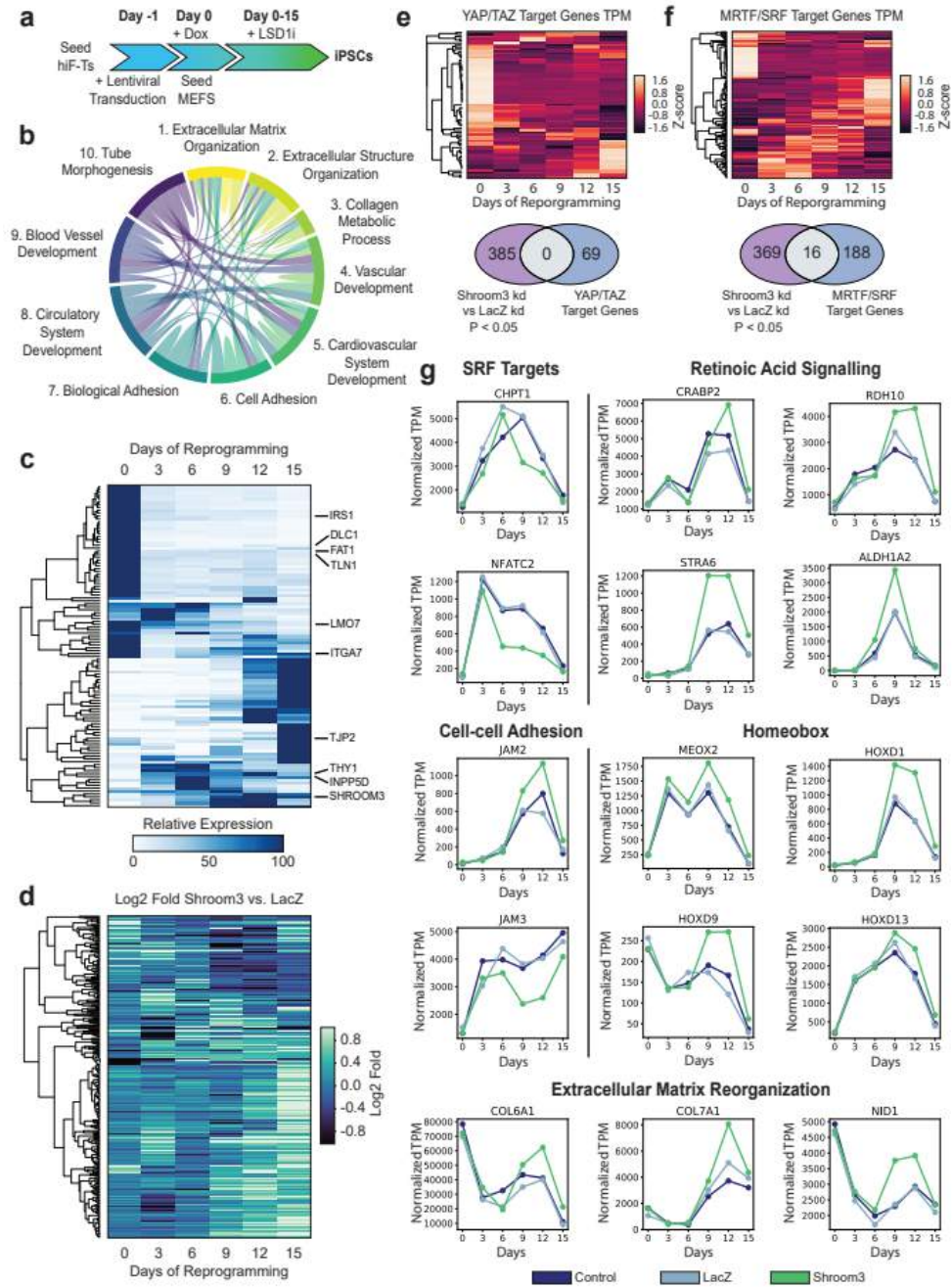

*Figure 21: SHROOM3 knockdown differentially regulates morphogenic and differentiation pathways (a) Experimental timeline for the SHROOM3 kd and LacZ non targeting control (b) Top 10 biological process gene ontologies enriched within genes differentially regulated by SHROOM3 kd (c) Adhesome gene expression in the control condition over the course of reprogramming (d) Heatmap depicting differentially regulated genes during reprogramming, scale reflects fold difference between SHROOM3 kd and LacZ non targeting control (e) YAP/TAZ target gene expression and overlap with differentially regulated genes (f) MRTF-A/SRF target gene expression and overlap with differentially regulated genes (e) Key differentially regulated gene expression profiles between SHROOM3, LacZ, and Control conditions.*

To answer these questions, we first narrowed the scope of our analysis to two well known mechano-sensitive pathways driven by the genes YAP/TAZ or MRTF-A/SRF respectfully[11,34]. In both cases, we found that the bulk of target genes downstream of each pathway are dynamically expressed during our reprogramming timeline. Out of the 385 differentially regulated genes affected by our SHROOM3 kd, none of them appeared to be downstream of YAP/TAZ (**Figure 21e**). However, our analysis revealed 16 genes downstream of MRTF-A/SRF are differentially regulated between our SHROOM3 kd and LacZ non-targeting shRNA control (**Figure 21e**). Out of these 16 genes, we saw a notable down-regulation of NFATC2 in the SHROOM3 kd condition as seen in (**Figure 21g**). NFATC2 is a transcription factor involved in non-canonical Wnt signalling[35], which is associated with ECM regulation, morphogenesis, and differentiation[29]. It is, therefore, possible that changes in apical constriction, due to a reduction of SHROOM3 at a protein level, results in decreased activation of MRTF-A/SRF which explains the observed down-regulation of NFATC2. This down-regulation could theoretically result in a decrease of NFATC2 associated differentiation pathways. However, the majority of differentially regulated genes do not at first appear to be downstream of MRTF-A/SRF.

Further analysis indicates some homeobox genes such as MEOX2, HOXD1, HOXD9, and HOXD13 are transiently up-regulated in the SHROOM3 kd relative to the non-targeting control. These genes are associated with morphogenesis and late embryogenesis. Intriguingly, the transient up-regulation of these genes corresponds to a similar increase in genes related to retinoic acid (RA) signaling and retinol metabolism, namely STRA6,

RDH10, CRAPB2, and ALDH1A2. Previous work indicates that retinoic acid signaling itself may be enough to activate the expression of HoxD cluster genes[36]. Previous work indicates that ROCK activation and signaling has the potential to down-regulate retinoic acid signaling by modulation of retinol metabolism[37]. Taken together with the knowledge that SHROOM3 is an activator of ROCK[32], it may be possible that the knockdown of SHROOM3 reduces ROCK activation, causing an increase in the expression of RA signaling and metabolism genes as well as specific homeobox genes. This may be critical for the progression of reprogramming.

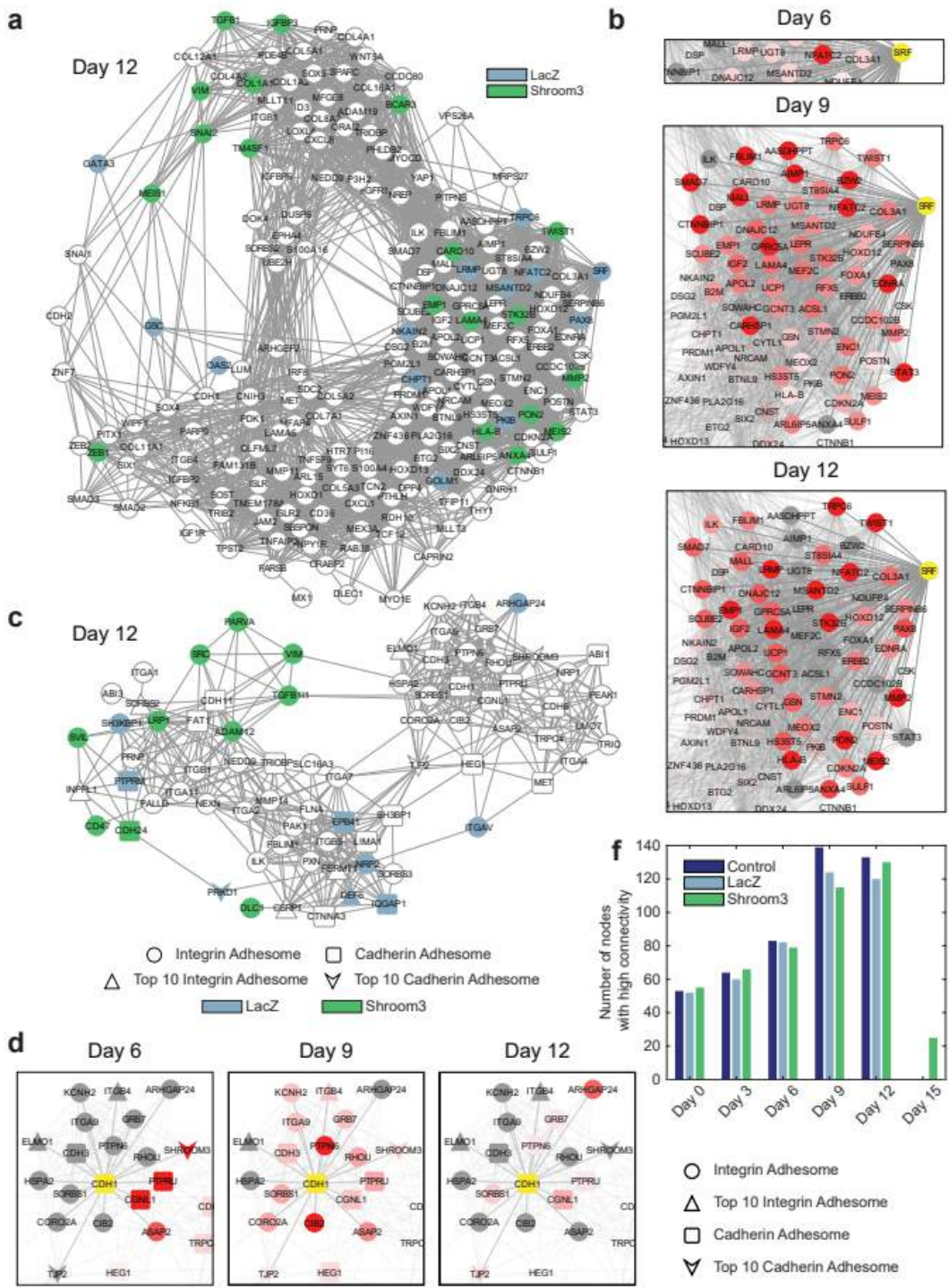

*Figure 22: Weighted gene correlation network analysis reveals a critical state transition regulated by SHROOM3 and SRF (a) General gene network on day 12 of reprogramming, built using genes differentially regulated between SHROOM3 and LacZ conditions (b) Correlation between SRF and other networked genes, redness indicates the strength of the correlation (c) General gene network on day 12 of reprogramming, built using Adhesome genes, Genes unique to SHROOM3 or LacZ networks are highlighted (d) Correlation between CDH1 and other networked genes (f) Network density throughout reprogramming, represented by the number of nodes with high connectivity, for all conditions.*

***Section 2.4: Gene Correlation Networks Reveal a Critical State Transition Modulated by SHROOM3 Gene Expression***

Next, we wanted to leverage our replicate time trials to identify or strengthen the case for other possible modes of action through which SHROOM3 could affect reprogramming. Previously work has suggested an unknown stochastic intermediate phase may be responsible for rate-limiting the process of reprogramming and thus reducing the overall reprogramming efficiency[4]. Furthermore, a transiently expressed gene network could theoretically orchestrate such a critical transition. Given SHROOM3's gene expression profile is transient during reprogramming, we hypothesized that it may influence a regulatory network of genes that may be in part responsible for the yet identified critical state transition required for the eventual induction of pluripotency. To test this hypothesis, we generated gene expression correlation networks for each condition at each time point (**Figure 22**). The genes, used to construct these networks, were transiently regulated in the control condition and differentially regulated between the kd conditions. Additionally, network construction included transcription factors, whose target sites were significantly enriched in the differentially regulated genes sets, and classical EMT genes.

We found that in all conditions, the networks we generated achieved peak connectivity on days 9 and 12 of reprogramming as represented by the number of highly connected nodes for each condition in (**Figure 22f**). By comparing the networks between the LacZ and SHROOM3 kd conditions, we were able to identify genes that were differentially networked between these conditions. Intriguingly, we found SRF was networked in the LacZ condition on multiple days but was absent from the SHROOM3 condition. SRF in the LacZ condition was found to network with NFATC2 and other WNT signalling genes, as might be expected. One explanation for these robust results is the known relationship between ROCK activation and SRF gene regulation[34]. Thus the kd of SHROOM3 may transiently reduce ROCK activation and subsequently SRF regulation of differentiation pathways during a critical transition within the reprogramming process.

# SECTION 3: Summary and Conclusions

***The Biophysical & Epigenetic Regulation of Macrophage Activation***

Concerning macrophage activation, these findings show a clear connection between morphological changes initiated by micropatterned topographies and the epigenetic regulation of proinflammatory and adhesome gene expression. Not only do micropatterned topographies elongate M1 macrophages, but they also reduce iNOS and CCL2 inflammatory gene expression and global H3Ac. Treating M1 macrophages with iBET achieves similar results. Such similarities indicate a possible regulatory overlap between the effects of iBET treatment and micropatterned surfaces, though further study is needed. By investigating gene expression changes caused by micropatterning in a broad set of inflammatory-related genes, we find that micropatterning reduces the expression of some inflammatory genes but increases others. However, iBET overwhelmingly reduces inflammation-related gene expression suggesting the mechanisms inherent to either condition are not entirely the same.

In order to get a better understanding of the pathways involved in the regulation of M1 macrophage behavior by micropatterning, we built a gene correlation network using genes differentially regulated by micropatterning and the adhesome. In this manner, we show that genes regulated by micropatterning are co-regulated along with a variety of integrin and cadherin related genes. Though further experimentation is necessary to infer causality. Furthermore, a differential reanalysis of previous data suggests a particular node in this network, composed of Cav1, ItgaV, and Nfκb, a fundamental inflammatory transcriptional regulator, are possibly co-regulated by changes in H3Ac within their gene promoters. This novel finding suggests there may be an epigenetic explanation for some of the gene expression changes observed when M1 macrophages are cultured on a micropatterned surface. A mechanistic study of such a relationship could prove fruitful.

Furthermore, the reduction of nuclear H3Ac in M1 macrophages caused by micropatterned surfaces can be recapitulated by treating M1 macrophages with various cytoskeletal inhibitors. This observation suggests macrophage cytoskeletal reorganization, caused by micropatterned surfaces, may ultimately result in the observed epigenetic and gene expression changes. Though further study is needed. Intriguingly, M1 macrophage motility increases in response to both iBET and micropatterned surfaces. These increases are further indication that macrophage morphology and adhesion are linked to the epigenetic regulation of M1 macrophage activation.

***Cell-generated force and adhesion contributes to bottleneck during cell reprogramming***

Concerning somatic cell reprogramming, our work indicates there is a bottleneck to reprogramming that is regulated by various adhesome genes and mechanosensitive pathways. Both ECM associated adhesion and Piezo1 agonist Yoda1 decrease the efficiency of the reprogramming process. Furthermore, cell generated forces peak on day 12 of reprogramming, which corresponds with the observed activation of Piezo1 associated calcium signaling. Together, with previous work from other researchers, it seems cellular force and adhesion mechanics play a pivotal role in the success of reprogramming. Furthermore, the transient up-regulation of cell generated forces indicates there are adhesion and mechanical related transitions during the reprogramming process that are worthy of further investigation.

By observing adhesome gene expression throughout the reprogramming process, we can safely say that there is a wide variety of unique ways in which reprogramming dynamically regulates adhesome gene expression. Intriguingly, a vast majority of these dynamically expressed adhesome genes appear to be barriers to reprogramming, as suggested by our shRNA knockdowns of these same genes. Furthermore, weighted gene correlation network analysis reveals that adhesome genes are highly co-regulated on day 12 of the reprogramming process. These findings suggest that the adhesome is not only regulated by various cell fate transitions but is an essential regulator of said transitions during the reprogramming process. Though mechanistic studies are necessary to prove such a regulatory relationship.

Next, we investigated the effect of the SHROOM3 kd, the top-performing shRNA knockdown, on gene expression dynamics during the reprogramming process. While knocking down SHROOM3 increases reprogramming efficiency by around 27-fold, it also differentially regulates a variety of genes associated with differentiation and morphogenesis. These included gene targets of SRF, a mechanosensitive transcriptional regulator. Furthermore, gene expression correlation analysis identified a likely critical state transition at day 12 of reprogramming that appears to be linked to SRF gene expression in the control condition but not in the SHROOM3 kd. These findings suggest that SHROOM3 contributes to the activation of mechanosensitive pathways that negatively affect the efficiency of the reprogramming process, thus confirming the importance of adhesome gene expression and mechanical signaling during reprogramming. Perturbing these proposed relationships may allow us to prove that they are biologically relevant.

# FUTURE WORK

In order to ascertain the importance of the cytoskeleton in integrating biophysical signals resulting in epigenetic and transcriptional changes, we must first observe cytoskeletally regulated processes such as traction force generation. While morphological observations and motility provide a good indication of cytoskeletal regulation, they can not replace direct observation of intracellular tension. With that said, we aim to directly measure forces generated through cytoskeletal actin-myosin contraction using traction force microscopy. Previous work has shown that M1 activated macrophages elicit less traction force than unstimulated macrophages. As such, we might expect M1 macrophages treated with iBET or cultured on micropatterns to elicit lower traction forces.

Furthermore, we aim to elucidate the causal relationship between micropatterned surfaces and the reduction in histone acetylation. Previous work has recognized that spatial confinement is a potent regulator of HDAC3, which is itself a regulator of LPS inducible gene expression.[15] Given this information, we believe that HDACs may also change their activity levels in response to our micropatterned surfaces. However, even if such an association is proven, the direct involvement of HDAC3 in the micropatterned induced differential expression of inflammatory genes remains unknown. As such, we would likely need to use genome-wide H3Ac ChIP-seq as a means of identifying differential H3Ac changes on micropatterned surfaces. In doing so, we may further elucidate the complicated dynamics between gene expression, histone acetylation, and micropatterns.

While our identification of differentially regulated genes and our subsequent network analysis suggest that our kd of SHROOM3 is perturbing a critical state transition in the reprogramming process, further analysis is needed. The proposed relationship between SHROOM3 expression and this network of genes could be strengthened by exogenously expressing SROOM3 in reprogramming cells and observing the transient expression of essential genes within the network. While a knockdown of SHROOM3 was shown to increase the expression of some of these genes transiently, we hypothesize that increased SHROOM3 expression would have the opposite effect. Furthermore, knocking out SHROOM3 expression altogether may give us a clearer understanding of its importance in regulating the reprogramming process.

Additionally, we do not know if the transient upregulation of genes identified within the network is a result of a shift in the heterogeneity of cell states during reprogramming or a result of gene upregulation within individual cells. Previous work has shown that gene expression within cells is different for those that are fated to reprogram successfully as opposed to unsuccessfully.[2] With that said, we propose using single-cell sequencing on cells at day 12 of reprogramming to determine if shifts in cell state heterogeneity explain our previous findings. Furthermore, the gene expression data gathered from single-cell sequencing may allow us to generate more insightful gene correlation networks as it accounts for bulk population dynamics.

Lastly, we aim to apply our methodology to study other cell fate transitions and in doing so, advance our understanding of potential regenerative therapies and cancer biology. Direct reprogramming is a technique uniquely poised to address injury in cardiac and nerve tissue.[38,39] However, the challenges associated with direct reprogramming remain similar to those experienced during the induction of pluripotency. By leveraging our techniques, developed for the study of somatic cell reprogramming, we hypothesize that we can similarly address bottlenecks experienced during direct reprogramming. Furthermore, we also believe that these techniques apply to other kinds of cell fate transitions, such as those in carcinogenesis, which is associated with changes in adhesome gene expression. [40]

# METHODS

## *Macrophage Culture and Activation*

Macrophages were derived and stimulated as previously described.[5–7] First mice were sacrificed and their tibia or femur extracted. Subsequently the bone marrow from each mouse was flushed out and treated with red blood cell lysis buffer. The remaining marrow derived cells were cultured in media with macrophage colony stimulating factor for 7 days. On day 3, all undifferentiated monocytes were washed away as only the differentiated macrophages remain. This population of cells was seeded at densities varying between 8,000 to 16,000 cells per cm$^2$ and allowed to adhere to the surface for 2 hours prior to stimulation. The macrophages were subsequently stimulated with 1ng/mL of INFγ and LPS for 2 or 16 hours resulting in M1 macrophage activation. In some cases cells were treated with 5 μM of iBET 30 minutes prior to and during stimulation. After stimulation, the cells were either fixed for immunostaining, the RNA was collected for qPCR and NanoString gene expression profiling, or proteins were collected for western blots.

## *Somatic Cell Reprogramming*

Reprogramming was carried out using an immortalized BJ fibroblast derived cell line known as hiF-Ts. These cells contain a doxycycline inducible OKSM polycistronic gene cassette as well as telomerase expressing vector as previously described.[2] These cells were expanded in growth media (GM) consisting of DMEM F12 with 10% embryonic stem cell grade fetal bovine serum (ES-FBS) in the presence of puromycin to ensure clonal purity. Prior to reprogramming puromycin is removed from the culture medium. On day -1 of reprogramming, the hiF-Ts were seeded at 75,000 cells per cm$^2$ and transduced with shRNA expressing lentiviral vectors as needed. On day 0 irradiated mouse embryonic fibroblasts (MEFs) are added at 150,000 cells per cm$^2$ and doxycycline was added to the GM in order to initiate reprogramming. Additionally on day 0, 1 µM of LSD1i and ROCKi are added to the GM as needed. On day 3, the 10% ES-FBS in the GM is replaced with 20% knockout serum replacement and the concentration of LSD1i is reduced to 0.1 µM for the duration of reprogramming.

**Microcontact Printing**

Microcontact printing was carried out as previously described.[7] All micropatterned surfaces presented here consisted of repeating fibronectin lines of 5 µm with 5 µm gaps in between. Cells were seeded on these surfaces

**shRNA Dynamic Adhesome Screening**

Adhesome gene targeting lentiviral shRNA constructs were drafted from The RNAi Consortium's library of shRNAs. 3 shRNAs per each 103 adhesome gene were tested on an individual basis. Viral titer concentrations were standardized prior to experimentation. Reprogramming was conducted with both ROCKi and LSD1i or neither. The iPSC colony counts were averaged for each gene targeting shRNA trio before fold changes were calculated. Conditions with outlying viral titer concentrations were discarded from the analysis. The best performing of each shRNA trio was selected for further experimentation as needed.

**Cellular Assays**

Reprogramming efficiency was assessed by staining and identifying TRA-1-60 positive colonies. First cells were fixed with 1% para-formaldehyde for 20 minutes, washed, and subsequently blocked and permeabilized for 30 minutes with a 3% bovine serum albumin and 0.02% Triton-X detergent. Next the samples were incubated with a biotinylated TRA-1-60 antibody at 4°C for 16 hours, washed and then incubated with a horse radish peroxidase conjugated streptavidin for 3 hours at room temperature. Finally, we used DAB Peroxidase substrate kit from Vector Laboratories (SK-4100) to darken the TRA-1-60 positive colonies. The wells were then digitally scanned and positive colonies were counted using ImageJ. Fold changes were calculated by taking the ratio of colony count over the experimental controls.

Immuno-fluorecense staining was carried out using the same fixation and permeabilization methods. However, a primary H3Ac antibody was used for a 16 hour incubation at 4°C. After 3 hours of incubation with a fluorescent secondary, cells were imaged. Cells were also stained with DAPI and phalloidin. DAPI stains were used to define the boundaries of the cellular nuclei for the purpose of measuring immuno-stained H3Ac fluorescence intensity. Phalloidin and phase contrast images were used to define cellular borders for cell aspect ration measurements. All image processing was accomplished in ImageJ.

**Transcriptomic Profiling and Network Analyses**

Reprogramming was conducted over 15 days using LSD1i but not ROCKi. Cells were singularized and collected using Accutase and then underwent MEF depletion using magnetic bead separation (Miltenyi Biotec) on days -1, 0, 3, 6, 9, 12, and 15. Experiments were performed in replicate. RNA was extracted from the remaining cells and scored for quality. Having passed the quality check, mature RNA was selected using poly-A enrichment followed by library preparation and single read 100 cycle Illumina sequencing conducted by the UCI Genomics High Throughput Facility. Reads were then trimmed using CutAdapt [41] and subsequently underwent quality control using FastQC. Following these per-processing steps, we used Salmon to quantify gene expression as transcripts per million (TPM) for each gene in hg38.[42] Next, these TPM values were normalized prior to the identification of differentially expressed or temporally regulated genes using either DEseq2 or ImpulseDE2 respectively. Both gene expression accuracy and fold change comparisons were validated using External RNA Controls Consortium (ERCC) spike in controls. Gene set enrichment analysis and gene promoter motif enrichment analysis was conducted using HOMER as previously described. [43] Additional analysis was performed using Python.

Gene network correlation analysis for the macrophages gene expression data was performed by first filtering out genes with low expression. Next, correlation coefficients for all pairwise arrangements were calculated for a predetermined list of adhesome and inflammatory genes. Connections with a correlation coefficient of less than 0.9 were excluded and the remaining connections were visualized using a custom python script. Any other gene correlation network were produced as previously described.[44,45] Differential network analysis was carried out using DyNet.[46]

**Traction Force Microscopy**

Polyacrylamide (PA) gel substrates were prepared with a modified procedure of previously published protocols[26,47]. Briefly, glass bottom dishes were functionalized with 0.1 M NaOH and (3-aminopropyl) triethoxysilane followed by glutaraldehyde treatment[47]. Top glass cover-slips were functionalized with Poly-D-Lysine (0.1 mg/mL) and a 1:800 dilution of red fluorescent micro-spheres (0.5 µm carboxylate-modified, Thermo Fisher Scientific) in water.[26]. A solution of 10% acrylamide and 0.1% bis-acrylamide was prepared. Polymerization was initiated with the addition of 1:100 tetramethylethylenediamine (TEMED) and 1:10 of a 10% ammonium per-sulfate (APS) solution. 20 μL was promptly pipetted onto the functionalized glass bottom dish and the functionalized top glass coverslip was placed on top. The dish was then turned upside down to minimize gravity effects that could cause fluorescent microspheres to polymerize lower into the substrate. After polymerization, fibronectin (20 μg/mL) was conjugated to the surface of gels with sulfo-SANPAH (Thermo Fisher Scientific) according to previous protocols[47].

Reprogramming and cell singularization was conducted as described for transcription profiling. The prepared PA gels were rinsed before cells were seeded at 3,000 cells/cm$^2$. Traction force microscopy imaging was performed as previously described[26]. To quantify traction forces ImageJ was used to register the unaligned images. Next, particle image velocimetry and fourier transform traction cytometry were performed as previously described[48]. A custom code was written in Python and IJ1 macro language to batch process the single cell traction forces. A custom code was written in R to perform statistical analysis on those results.

# APPENDIX A

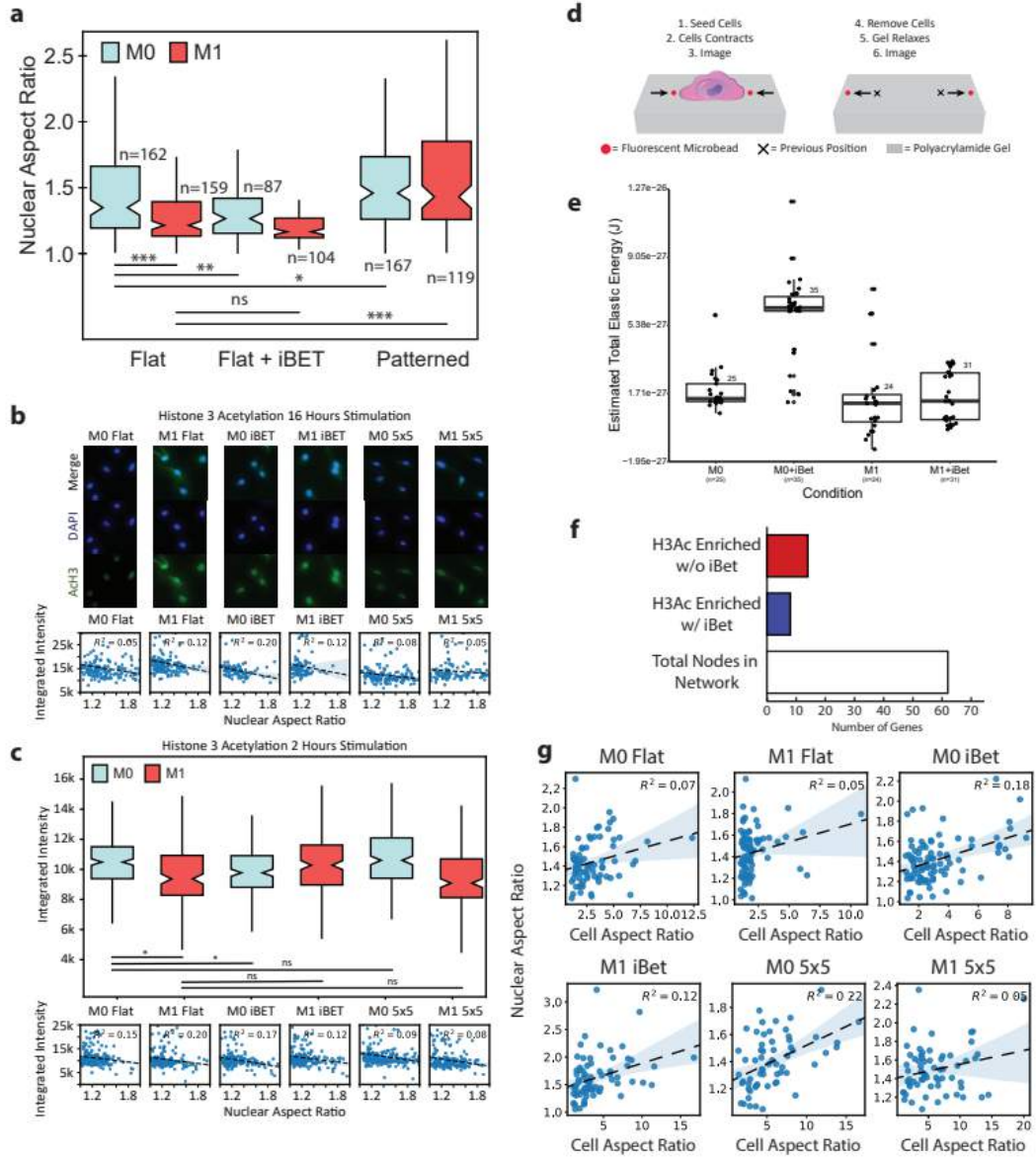

*Appendix 1: (a) Nuclear aspect ratio across all conditions. (b) H3Ac representative images and scatter plots between H3Ac fluorescent intensity and nuclear aspect ratio showing little to no correlation (c) H3Ac integrated immuno-fluorescence showing at 2 hours of stimulation showing little to no change across conditions (d) Representative schematic of traction force microscopy (e) Traction force microscopy for M0 and M1 macrophages with and without iBET (f) Number of genes, used to build the gene correlation networks, where H3Ac increases in response to LPS treatment exclusively with or without iBET treatment (g) Scatter plots between cell aspect and nuclear aspect ratio showing a weak positive correlation*

| Gene Name | T1 2hr 5x5 vs Flat Log2 | T2 2hr 5x5 vs Flat Log2 | Gene Name | T1 16hr 5x5 vs Flat Log2 | T2 16hr 5x5 vs Flat Log2 |
|---|---|---|---|---|---|
| Il12b | -2.717293658 | -1.905560955 | Nos2 | -1.086969359 | -0.84075732 |
| Ifit2 | -1.997973195 | 0.00385602 | Cxcl2 | -0.703337176 | -0.542154351 |
| Ccr1 | -0.819932198 | -0.831877241 | Ccl2 | -0.82922696 | -0.276953342 |
| Nfe2l2 | -1.299286415 | -0.092805164 | H2-Eb1 | -0.491926184 | -0.338390991 |
| Gnas | -1.142241443 | -0.028858579 | Cxcl3 | -0.495548324 | -0.078482515 |
| Ifit3 | -1.186463906 | 0.070798101 | Cfb | -0.510057447 | 0.046214108 |
| Il6 | -1.088805788 | -0.009533048 | Il1b | -0.496134703 | 0.176616007 |
| Cd86 | -1.180092319 | 0.166134379 | Tyrobp | -0.317018438 | 0.091826714 |
| Gnaq | -0.891164203 | 0.045143955 | Nlrp3 | -0.262589884 | 0.043596529 |
| Il1rn | -0.764921244 | -0.06593427 | Il6 | -0.615852363 | 0.463670088 |
| Mrc1 | -0.567803831 | -0.15181474 | Il1a | -0.430562423 | 0.429181545 |
| Stat3 | -0.973677333 | 0.283255325 | Tgfb1 | -0.166488327 | 0.19217033 |
| Iigp1 | -0.329327812 | -0.263756379 | Iigp1 | -0.083151915 | 0.140550188 |
| Ifit1 | -0.503109399 | -0.065281759 | Stat3 | -0.121161886 | 0.184309074 |
| Il15 | 0.012900489 | -0.485442533 | Cxcl9 | -0.134221185 | 0.203498943 |
| Oas1a | -0.309800616 | -0.058800376 | Stat1 | -0.224919033 | 0.307590553 |
| Rac1 | -0.163789305 | -0.152153712 | Hif1a | -0.193271797 | 0.304993862 |
| Cxcl9 | 0.061493615 | -0.332561788 | Cxcl10 | -0.108510326 | 0.225784848 |
| Nr3c1 | -0.468705584 | 0.207696935 | Ptgs2 | -0.034405636 | 0.166003878 |
| H2-Eb1 | -0.071272437 | -0.141388543 | Cfl1 | -0.093313496 | 0.258726426 |
| Itgb2 | -0.082009787 | -0.081605576 | Itgb2 | -0.103206566 | 0.286816082 |
| Ccl5 | -0.123126931 | -0.004504584 | Mapk3 | -0.101382011 | 0.397026707 |
| Cd40 | -0.128897482 | 0.008798114 | Ccl5 | -0.153569371 | 0.488996216 |
| Cxcl10 | -0.080628339 | 0.004350418 | Cdc42 | 0.078131379 | 0.279026264 |
| C1qa | 0.005807696 | -0.077497836 | Rac1 | -0.006702544 | 0.365251695 |
| Stat1 | 0.25689338 | -0.317984427 | C1qa | -0.219847589 | 0.629716774 |
| Csf1 | -0.072067483 | 0.060068347 | Nfe2l2 | -0.052869046 | 0.464903312 |
| Tlr2 | 0.135501857 | -0.107178987 | Rhoa | 0.089942889 | 0.340856761 |
| C3 | -0.048338467 | 0.117896363 | Map2k1 | 0.03759236 | 0.430654178 |
| Cfl1 | 0.036579824 | 0.033053733 | Gnb1 | 0.094251094 | 0.394213925 |
| Mapk3 | -0.04981282 | 0.13286859 | Nr3c1 | 0.095548949 | 0.403492263 |
| Trem2 | 0.006685461 | 0.076776421 | Myd88 | -0.037716289 | 0.550151172 |
| Il10rb | 0.070291166 | 0.044793402 | Cebpb | 0.025881211 | 0.506382426 |
| Ccl2 | 0.078923882 | 0.047303993 | C3 | 0.020084246 | 0.538093243 |
| Myd88 | 0.423759036 | -0.262704133 | C1qb | 0.062219763 | 0.511094526 |
| Tgfb1 | 0.174088248 | 0.019792952 | Cd40 | 0.057818448 | 0.520401272 |
| Irf1 | 0.17491924 | 0.021343856 | Irf1 | 0.136308871 | 0.64288226 |
| Cxcl2 | -0.243911681 | 0.492278339 | Gnas | 0.067617935 | 0.739004292 |
| C1qb | 0.310035599 | -0.034125557 | Relb | 0.275882786 | 0.642466702 |
| Tnf | -0.061006737 | 0.350404677 | Mapkapk2 | 0.332522823 | 0.612392337 |
| Ccl7 | 0.555318324 | -0.240934526 | Tnf | 0.252432699 | 0.708051749 |
| Cdc42 | 0.207615437 | 0.114343073 | Grb2 | 0.398534237 | 0.571751552 |
| Cxcl1 | 0.098455589 | 0.257442312 | Il10rb | 0.252468727 | 0.823944658 |
| Nlrp3 | 0.374634376 | 0.024037966 | Rapgef2 | 0.342955601 | 0.870589989 |
| Irf5 | 0.543516335 | -0.142891545 | Oas1a | 0.229670184 | 1.038200864 |
| Ifi27l2a | 0.403850457 | -0.001466088 | Nfkb1 | 0.174306645 | 1.106817431 |
| Hif1a | 0.331466528 | 0.08373609 | Il1rn | 0.629719079 | 0.705042336 |
| Tollip | 0.419562098 | 0.009114405 | Irf5 | 0.64467493 | 0.730679769 |
| Tnfaip3 | 0.176664825 | 0.260561733 | Stat2 | 0.427434595 | 0.982094598 |
| Rhoa | 0.546014271 | -0.097644557 | Ifi44 | 0.234968242 | 1.274159803 |
| Il1b | 0.148789829 | 0.301566928 | Cd86 | 0.347880791 | 1.204564804 |
| Tyrobp | 0.548867112 | -0.089351503 | Ccl3 | 0.5553098 | 1.136344843 |
| Cxcl3 | 0.183008153 | 0.297682548 | Oasl1 | 0.570864745 | 1.219238403 |
| Map2k1 | 0.138833952 | 0.35742725 | Ifit1 | 0.511565881 | 1.295451569 |
| Relb | 0.282804141 | 0.223533117 | Ifit2 | 0.405123792 | 1.40496013 |
| Cebpb | 0.543755846 | -0.013379304 | Irf7 | 0.593328444 | 1.222937322 |
| C3ar1 | 0.355722701 | 0.184870868 | Ifi27l2a | 0.576071435 | 1.250008754 |
| Pdgfa | 0.29927491 | 0.301565721 | Tnfaip3 | 0.652022589 | 1.199505992 |
| Rapgef2 | 0.583603042 | 0.04739805 | C3ar1 | 0.654381537 | 1.240064828 |
| Il1a | 0.249109301 | 0.497044433 | Mx1 | 0.722120809 | 1.580452168 |
| Gnb1 | 0.795941744 | 0.061861568 | Ifit3 | 0.721287052 | 1.608212279 |
| Nfkb1 | 0.430778869 | 0.448920507 | Trem2 | 0.897082805 | 1.503418647 |
| Oasl1 | 0.715086709 | 0.270151951 | Fos | 0.781251825 | 1.854015314 |
| Ptgs2 | 0.428211001 | 0.78839533 | | | |
| Mapkapk2 | 0.778647175 | 0.503082116 | | | |
| Ccl4 | 0.837717578 | 0.67200998 | | | |
| Ccl3 | 0.794473303 | 1.067001297 | | | |

*Appendix 2: Log2 Fold change gene expression Nanostring data for M1 macrophages on patterned surfaces relative to flat surfaces after 2 or 16 hours of stimulation. T1 and T2 represent biologically independent replicates. All genes displayed were expressed above a background threshold*

# APPENDIX B

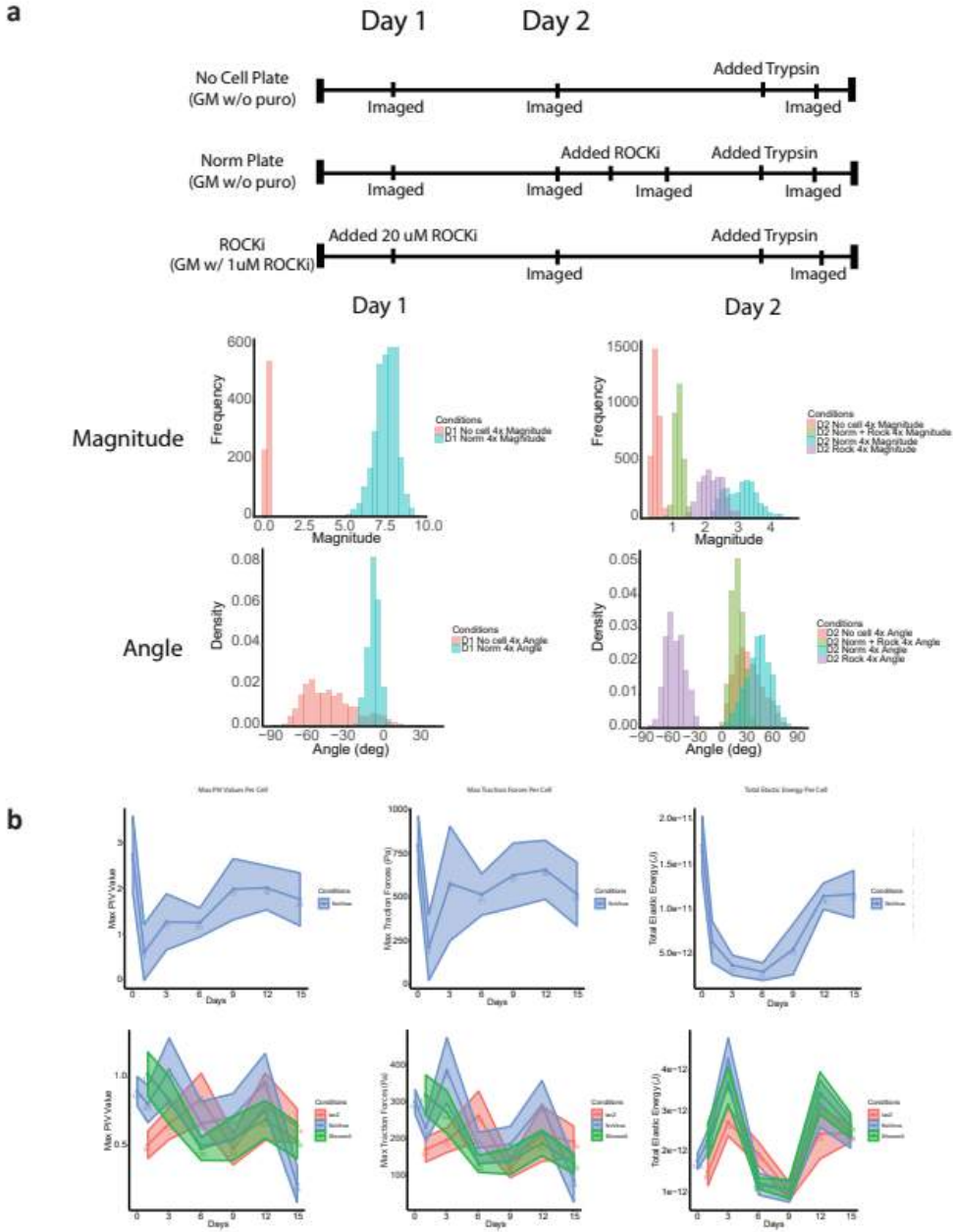

*Appendix 3: Traction force microscopy validation and reprogramming timeline(a) Validation of TFM showing a measurable difference between the distribution of gel deformation measurements under cell free, hiF-Ts, and hiF-Ts treated with ROCKi. (b) TFM timeline for normal reprogramming and reprogramming with shRNA kd of Shroom3 with controls. Substrate deformation and traction forces are shown as particle image velocimetry (PIV), maximum force for each cell, and total elastic energy exerted by each cell.*

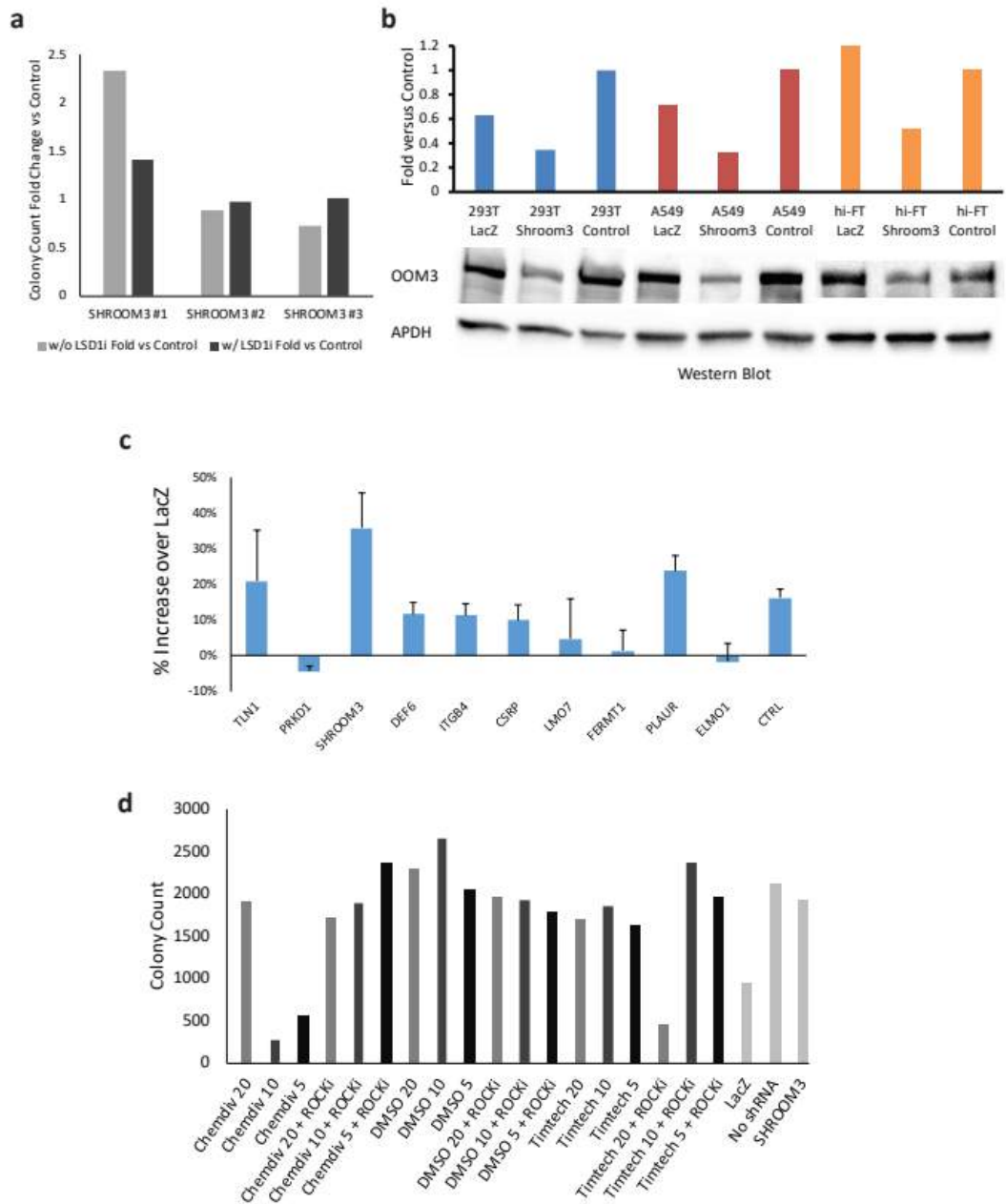

*Appendix 4: Validation of Top shRNA hits and SHROOM3 protein level knockdown.(a) shRNA kd of SHROOM3 targeting 3 separate sequences within the reference transcript. shRNA #1 improves reprogramming the most and was chosen for further study. (b) Western blot illustrating kd of SHROOM3 versus controls at a protein level in cell types 293Ts, A549s, and hiF-Ts when normalized to GAPDH (c) shRNA knockdown of 8 out of 10 successfully improved reprogramming over a non-targeting control when reprogrammed with LSD1i and not ROCKi. (d) Inhibition of Shroom3 using small molecules from Timtech or Chemdiv at different concentrations (μM) did not improve reprogramming against appropriate DMSO controls.*

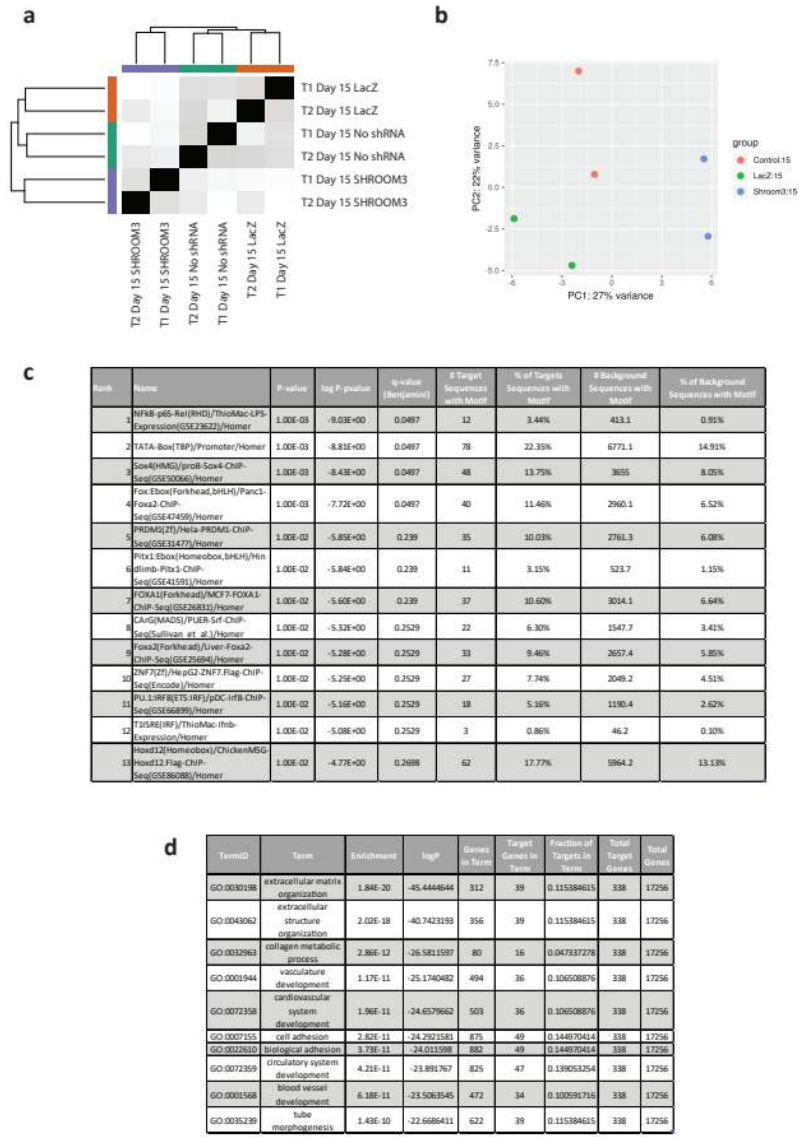

*Appendix 5: Similarity metrics and principle component analysis (PCA) reveals transcriptomic differences between SHROOM3 kd, non targeting controls, and no shRNA controls(a)(b) Correlation matrix, hierarchical clustering, and PCA of SHROOM3 kd and controls on day 15 of reprogramming. (c) Top 13 motifs enriched using HOMER to analyze the promoters of all differentially regulated genes between SHROOM3 kd and control during reprogramming. (d) Top 10 biological processes enriched using HOMER on all differentially regulated genes between SHROOM3 kd and control during reprogramming.*

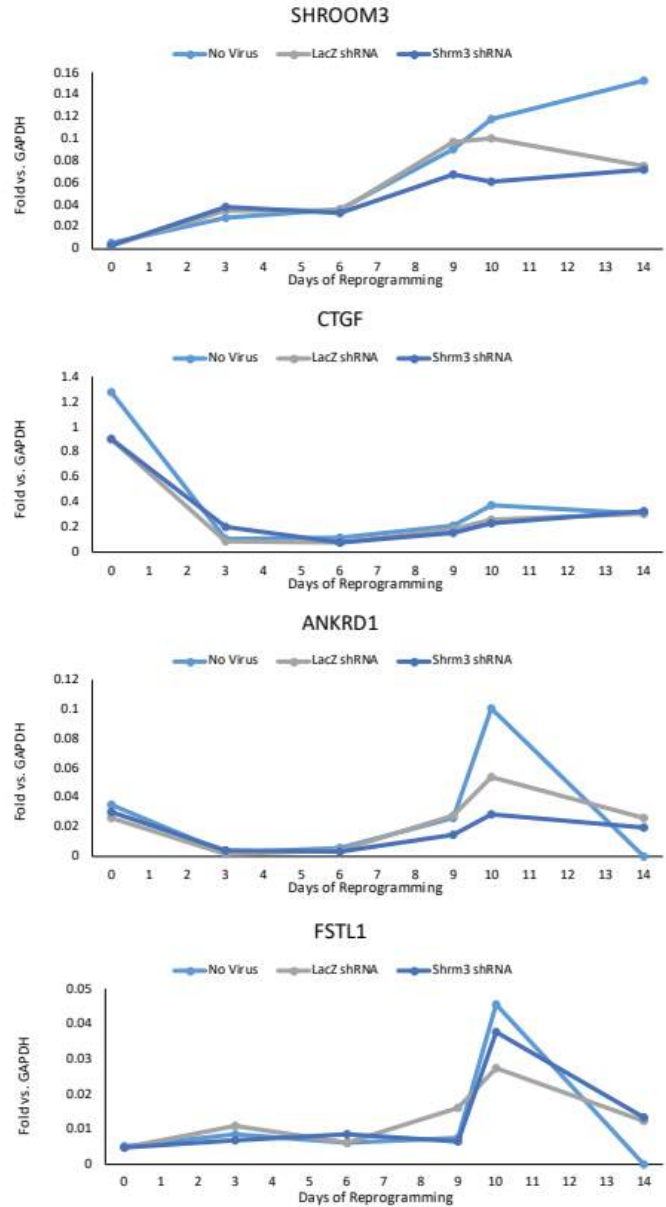

*Appendix 6: qPCR reveals shRNA kd of SHROOM3 reduces it's expression during reprogramming and may impact YAP/TAZ target expression(a) qPCR fold change in estimated expression against GAPDH for SHROOM3 and well known YAP/TAZ target genes CTGF, ANKRD1, and FSTL1.*

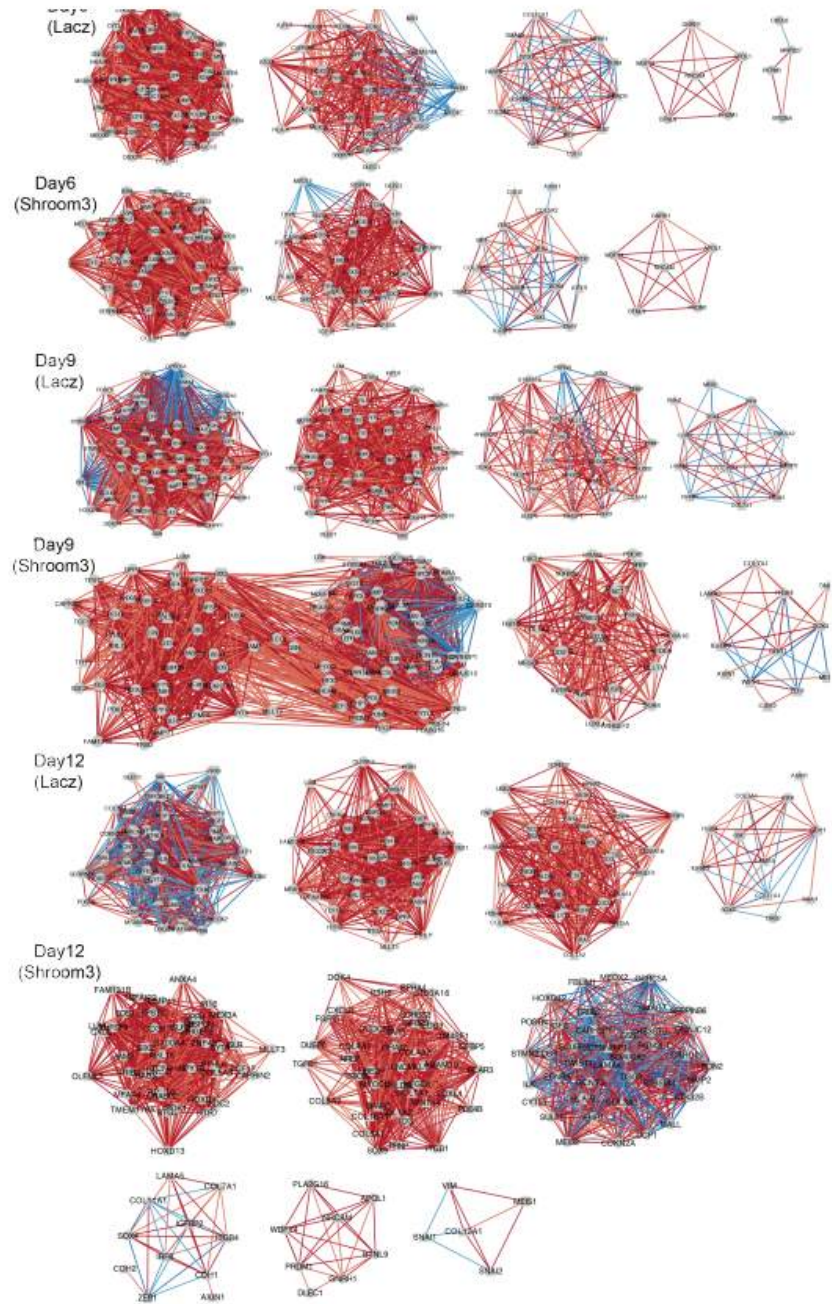

*Appendix 7: Clustered correlation networks reveal differential rewiring and possible regulatory networks between SHROOM3 kd and non targeting controls(a) Correlation networks for SHROOM3 kd and LacZ not targeting controls on days 6, 9, and 12 of reprogramming. Networks have been clustered into separate nodules based on network connectivity within each condition. Links in blue and red represent negative and positive correlation respectively.*

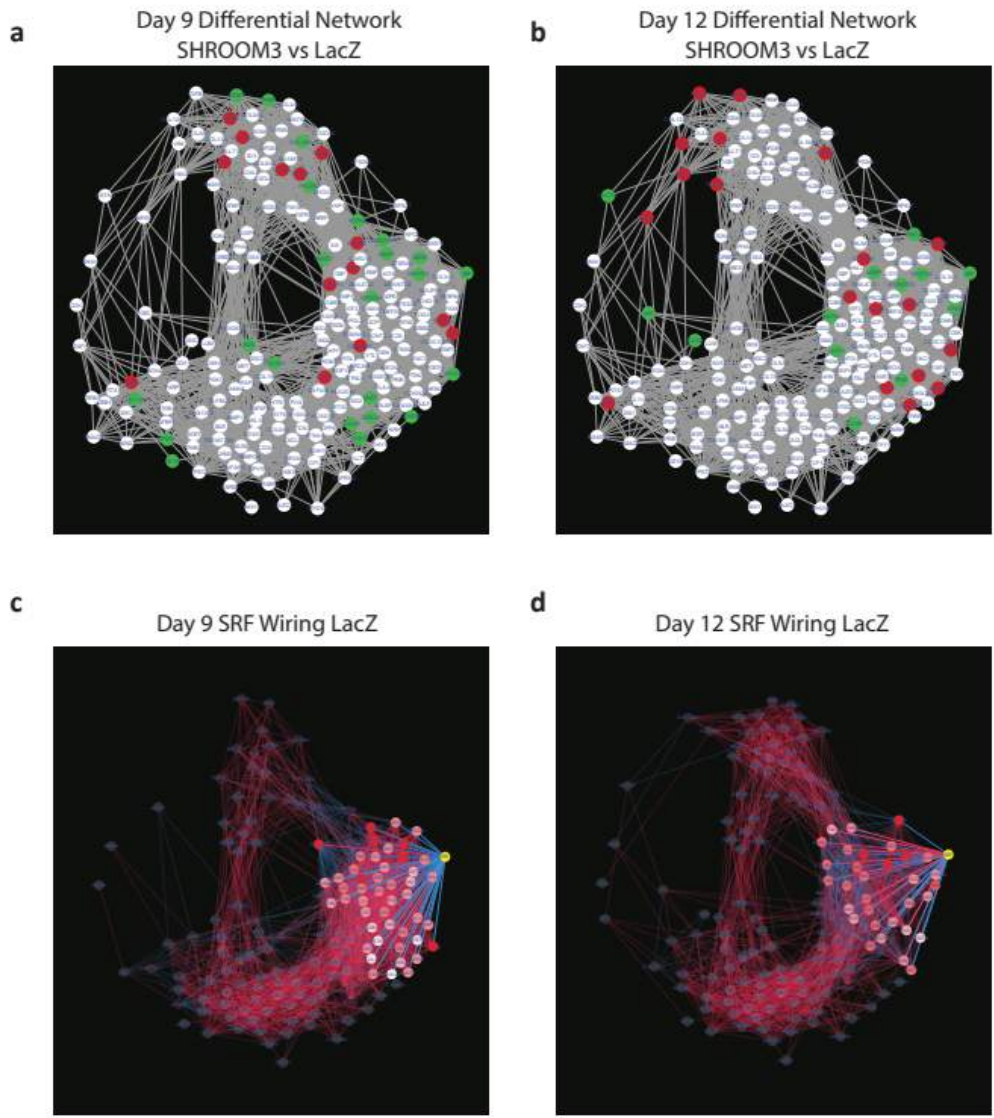

*Appendix 8: Differential comparison between correlation networks of SHROOM3 kd and non targeting controls illustrates possible regulatory differences between conditions(a)(b) Differential network comparing SHROOM3 and LacZ non targeting controls on days 9 and 12 of reprogramming respectively. Red and green nodes/genes indicate those unique to the SHROOM3 kd or LacZ non targeting control respectively. (c)(d) Networks indicating the correlation of the mechano-sensitive transcription factor SRF, unique to the LacZ correlation networks, on day 9 and 12 of reprogramming. Node color intensity represents strength of correlation with SRF expression. Links in blue and red represent negative and positive correlation respectively.*